# Nonequilibrium stochastic thermodynamics of boundary functionals: From Zubarev's ensemble to stochastic particle separation

V. V. Ryazanov, https://orcid.org/0000-0002-5308-3212

Institute for Nuclear Research, pr. Nauki, 47 Kiev, Ukraine, e-mail: vryazan19@gmail.com

This paper develops a generalization of Zubarev's nonequilibrium statistical operator method for the case of simultaneous inclusion of additive and nonlocal boundary functionals of trajectories. Using a unified thermodynamic approach, a three-parameter model of nonequilibrium systems is constructed, including the first-passage time, the dwell time above a given level, and the absolute extremum of the process. A correspondence is demonstrated between the maximum information entropy method for trajectories and the Donsker-Varadan large deviation formalism. Using the Doob`s h-transform, it is demonstrated that fixing extremal functionals of history induces efficient non-Markovian transport in the system with dynamic adaptation to historical records. Criteria for the applicability of the developed apparatus in the "large time" domain and the possibilities of its use for optimizing stochastic particle separation in periodic potentials are discussed.

## 1. Introduction

A fundamental problem in nonequilibrium statistical mechanics is the construction of a closed macroscopic description of open stochastic systems subject to intense thermal fluctuations. The traditional thermodynamic formalism, rooted in the work of Gibbs and Onsager, operates with ensembles of instantaneous values of phase variables, such as energy, position, or particle number. In this classical approach, rapid microscopic fluctuations are averaged out, and the evolution of the probability density in phase space is considered a strictly memoryless Markov process.

However, in a wide class of physical problems—including the dynamics of macromolecules and polymer chains in nanopores, transport in biological motors, glass transition processes, and the kinetics of metastable phase transitions—the macroscopic nonequilibrium response of a system critically depends not on the instantaneous state of the parameters, but on the intensity and geometry of the fluctuation history along the entire trajectory. Attempts to describe such systems within the framework of standard Markovian Fokker–Planck equations inevitably lead to the loss of non-Markovian memory effects or require the introduction of non-local integro-differential operators such as Caputo, which are mathematically cumbersome and preclude the direct construction of thermodynamic potentials.

Stochastic Thermodynamics approaches, the theorems of Jarzynski, Crookes, Refs. [1-5] work with trajectories and times, but are focused on microscopic single molecules (e.g. stretching DNA with laser tweezers). Jarzynski's theorems link work and free energy along trajectories. However, they do not connect the mesoscopic and macroscopic levels (Continuum Mechanics), allowing fluctuating time parameters to be included directly in the equations of continuous media (hydrodynamics, thermal conductivity), nor do they serve as a "bridge" (interface) between the microworld of single molecules and the macroworld of classical continuum physics.

In complex physical processes (polymer chains, metastable states, biological motors), the macroscopic response critically depends not on instantaneous coordinates, but on the history of fluctuations along the trajectory. The standard Markov formalism averages these fluctuations, losing non-Markovian effects ("memory"). In standard statistical physics, the Fokker-Planck Markov operator describes the current state of a system. The probability of a system transitioning to the next state depends only on its current location, while its entire past is erased by averaging.

A rigorous physical method for solving this problem is based on the nonequilibrium statistical operator (NSO) method by D. N. Zubarev Refs. [6-10]. The NSO method, based on the Shannon-Gibbs principle of maximum information entropy under given macroscopic constraints, allows one to naturally include the prehistory of a system in a statistical ensemble. In Refs. [11], the high efficiency of applying boundary functionals of random processes in statistical physics was demonstrated. In developing this concept, in Refs. [12-16], the boundary functional of the first-passage time (FPT) was used as a thermodynamic order parameter, based on the distribution of which a closed nonequilibrium FPT thermodynamics was constructed. The inclusion of boundary functionals in the description as additional thermodynamic variables corresponds to the formulation of the problem of multiparameter thermodynamics of the fluctuation history in the context of Refs. [11] and Refs. [12-16].

Despite the success of single-parameter FPT models, they suffer from a fundamental limitation: by averaging the system over all other trajectory descriptors, they are unable to separate the effects of the rate of reaching the critical state from the duration of its retention. This work aims to overcome this limitation by moving to multiparameter nonequilibrium thermodynamics of fluctuation history.

In this paper, we propose to extend Zubarev's thermodynamic basis by simultaneously introducing three fundamentally different classes of trajectory functionals: the additive residence time of a process above a given level $g_a$, the boundary first-passage time $\tau_{\mathbf{FPT}}$, and the nonlocal absolute extremum of a trajectory $M_\tau$. We show that the maximum information entropy method for this extended basis is equivalent to the strict Escher shift of measures in the theory of random processes Ref. [17] and the Donsker–Varadan theory of large deviations Ref. [18]. The residence time functional allows a rigorous description of the kinetics of rare large deviations and metastability without complicating the original microscopic equations. The introduction of a multiparameter description expands the boundaries of the thermodynamic approach, adding the capabilities of rigorous stochastic methods.

By expanding a closed system of equations in an extended phase space, we demonstrate how conjugate thermodynamic history fields replace integro-differential memory tails, collapsing information about the past into the geometry of the Doob's effective information potential. Within the framework of a linear response, a symmetric matrix of mutual Onsager susceptibilities is derived, capturing hidden cross-correlations of history. In the nonlinear limit of strong fields, the effect of self-similar trajectory confinement and the generation of singular transport with a memory of records are discovered. Finally, we demonstrate the effectiveness of the developed apparatus by solving the problem of optimizing stochastic particle separation in periodic potentials, where phase separation is achieved not through macroscopic mobility, but through subtle differences in the topology of the fluctuation history.

The paper is organized as follows. Section 2 introduces the generalized partition function and discusses its relationship to large deviation theory. Section 3 presents the dynamic equations for the trajectory ensemble. Section 4 considers the linear response regime. The nonlinear limit of strong fields and self-similar asymptotics are discussed in Section 5. The results obtained in Section 6 are applied to the optimization of stochastic particle separation. Section 7 discusses the results and criteria for the applicability of the theory. Section 8 contains the conclusion and conceptual implications.

## 2. Extended Zubarev trajectory ensemble and probabilistic formalism

### 2.1 Microscopic dynamics and trajectory functional

Microscopically, the initial dynamics of the system are defined by the standard Markov Langevin equation in the static potential relief U(x). However, the Zubarev procedure of selecting trajectory functionals leads to a dual macroscopic description. We consider the stochastic dynamics of the system, defined by the multidimensional Langevin equation:

$$\dot{\mathbf{x}}(t) = \mathbf{F}(\mathbf{x}) / \gamma + \sqrt{2D}\xi(t), \quad \langle \xi_i(t)\xi_j(t^{'})\rangle = \delta_{ij}\delta(t-t^{'}), \tag{1}$$

where $\mathbf{F}(\mathbf{x}) = -\nabla U(\mathbf{x})$ is the external potential field, $\gamma$ is the friction coefficient, $\mu(x) = \mathbf{F}(\mathbf{x}) / \gamma$ is the drift, D is the diffusion coefficient, and $\xi(t)$ is white noise. It corresponds to the standard Fokker-Planck operator L: $L\rho = -\frac{\partial}{\partial x}\left[\mu(x)\rho\right] + D\frac{\partial^2 \rho}{\partial x^2}$. Let's move on to the modified Fokker-Planck equation.

To correctly introduce the conjugate field $\lambda_\Gamma$ into the evolution equations, we need to use the formalism of modified (tilt) evolution operators, which is actively used in large-scale deviations (Large Deviation Theory Ref. [18]) and stochastic thermodynamics Refs. [1-5]. The field $\lambda_\Gamma$ is conjugate to the trajectory functional of the system's residence time in the subspace $\Omega_a = \{\mathbf{x} : f(\mathbf{x}) \geq a\}$ over the macroscopic time interval [0, $\tau$ ], which is introduced as:

$$\hat{\Gamma}_a = \int_0^\tau \Theta\big(f(\mathbf{x}(t)) - a\big)dt , \tag{2}$$

where $\Theta(z)$ is the Heaviside function $\Theta(z)=1$ for $z \geq 0$ and $\Theta(z)=0$ for $z < 0$. We introduce a boundary or trajectory functional of the form $\hat{F}_\Gamma = \int_0^\tau f(x(t))dt$ (for an additive functional) or take it into account through boundary conditions (for the time of first reaching FPT; $\tau^+(x) = \inf\{t : \xi(t) > x\},\ x > 0$ is the moment of the first exit for the level x>0). Introducing a conjugate field $\lambda_\Gamma$ in the spirit of Zubarev's method is equivalent to switching to a modified probability density (biased or tilted distribution) $\tilde{\rho}(x,t) = \left\langle \delta(x(t)-x)e^{-\lambda_\Gamma \hat{F}_\Gamma} \right\rangle$.

The modified direct Fokker-Planck equation is:

$$\frac{\partial \tilde{\rho}(x,t)}{\partial t} = L\tilde{\rho}(x,t) - \lambda_\Gamma f(x)\tilde{\rho}(x,t) . \tag{3}$$

The physical meaning of expression (3). The conjugate field $\lambda_\Gamma$ generates an additional "sink-source" potential (absorption or propagation of trajectories) $-\lambda_\Gamma f(x)$. The system no longer preserves the total probability, since the evolution operator becomes non-Hermitian and does not preserve the normalization. The evolution of the normalization determines the generalized Zubarev partition function Q(t).

Let's write a modified inverse Kolmogorov equation. For the conjugate function (initial state generator) $q\left(x_0,\ t\right)$, the equation takes the form:

$\dfrac{\partial q(x_0,t)}{\partial t} = L^\dagger q(x_0,t) - \lambda_\Gamma f(x_0) q(x_0,t)$, where $L^\dagger = \mu(x_0)\dfrac{\partial}{\partial x_0} + D\dfrac{\partial^2}{\partial x_0^2}$ is conjugate operator.

If the functional is the time of first reaching $\tau_{FPT}$, then the term $-\lambda_\Gamma f(x)$ disappears from the equation itself, but the field $\lambda_\Gamma$ "goes" to the boundary condition: at the absorbing boundary *a* the distribution function is zeroed, and the conjugate operator generates an exponential shift of the spectrum $L^\dagger$ by the amount $\lambda_\Gamma$.

Here, the Langevin approach (1) (trajectory/Heisenberg) is presented. We track an individual "particle" (or point in phase space). The dynamics are defined by a stochastic differential equation, where fluctuations are represented by a random force. Boundary functionals of the form (2) are formulated here naturally, since they are calculated directly from the trajectory X(t).

The Liouville/Fokker-Planck approach (3) (field/Schrödinger). We abandon tracking a single trajectory and look at the evolution of the probability density $\rho(x,t)$ over the entire phase space. This is a deterministic partial differential equation.

How do they combine when introducing boundary functionals? To find the distribution of a boundary functional (formulated in Langevin terms) through the Liouville/Fokker-Planck equation, physicists use absorbing or reflecting boundary conditions.

For example, for the first-hit time (FPT). 1. In Langevin terms: we seek the instant τ when the trajectory X(t) first crosses the boundary *a*. 2. In Fokker-Planck terms: we solve the probability density equation, but impose a zero (absorbing) boundary condition at the boundary *a*. The decay of the total probability from the volume with time gives us the FPT distribution.

Thus, adding a boundary functional to Zubarev's method is a way to transfer Langevin's trajectory information to the language of the evolution of the distribution operator (Liouville), modifying its structure.

The Langevin equation for the potential landscape, which is used in trajectory thermodynamics Refs. [19-20] and Zubarev's method, is microscopically the usual Langevin equation, but its macroscopic ensemble is structured differently.

Here an important physical duality arises between the original equation (in which a single particle moves) and the effective equation (which is seen by an observer fixing the boundary functionals).

1. Microscopic level: This is a common equation. At the level of a single isolated trajectory x(t), this is the most standard classical Langevin equation (1) (in the highly viscous approximation): $\dot{x}(t)=\frac{F(x)}{\gamma}+\sqrt{2D}\xi(t)=-\frac{1}{\gamma}\frac{\partial U(x)}{\partial x}+\sqrt{2D}\xi(t)$, where U(x) is a static, time-independent potential landscape (e.g., a periodic sine wave or parabola of an oscillator), and $\xi(t)$ is standard white Gaussian noise. This formula has no memory, no nonlocality, and no fractional derivatives. The particle simply rolls along the terrain U(x) under the impact of the thermostat.

2. The macroscopic level corresponds to the transition to a "not quite ordinary" ensemble. What makes this formalism "not quite ordinary" is the procedure of statistical selection of trajectories (the s-ensemble Refs. [19-20] or the Zubarev ensemble Refs. [6-10]).

When we require that the mean value of a history functional (e.g., the sojourn time $g_a$) be fixed, we take an infinite set of solutions of this ordinary Langevin equation and weight each trajectory by the Escher factor $e^{-\lambda_a\Gamma_a}$ (6) Ref. [17].

From the point of view of the theory of random processes (Girsanov's theorem Ref. [21]) and stochastic thermodynamics, this selection of trajectories is equivalent to the fact that our particle begins to obey a new, effective Langevin equation (see section 3.2):

$$\dot{x}(t)=v_{eff}(x)+\sqrt{2D}\xi(t)=\left[-\frac{1}{\gamma}\frac{\partial U(x)}{\partial x}+2D\frac{\partial\ln\phi_0(x;\lambda_a)}{\partial x}\right]+\sqrt{2D}\xi(t), \tag{4}$$

where $\phi_0$ is the main eigenfunction of the Feynman-Katz equation (9) Ref. [22].

What are the key differences between the effective dynamics and the "ordinary" ones? In the ordinary Langevin equation, the particle will wander naturally. But to model a system with a fixed history (field $\lambda_a$ ), you have to integrate the effective equation, which has three fundamental features: 1. The emergence of an information potential. The particle now moves not in the original potential U(x), but in the Doob's effective potential $[U(x)-2\gamma D\ln\phi_0(x)]$ (10) Ref. [23]. The "history field" $\lambda_a$ completes the geometry of the landscape, curving it so that rare trajectories become typical (as in Fig. 5, where the barriers disappeared for one type of particle, and deepened for another). 2. Explicit non-Markovian property (for extremes). If the parameter base contains a maximum $M_\tau$, then the effective drift begins to depend on the current record m: $v_{eff}=v_{eff}\left(x,\ m\right)$. The dynamics of a single particle becomes non-Markovian - the force pushing it at a given moment "knows" how high the particle flew in the past (Fig. 1). 3. Violation of the fluctuation-dissipation relation (FDR). In the conventional Langevin equation, the diffusion coefficient D and friction $\gamma$ are rigidly related by the thermostat temperature T (Einstein's relation $D=k_BT/\gamma$ ). In the effective Doob's equation, the restoring force $2D\partial_x\ln\phi_0$ is generated not by the thermostat, but by the information constraint of history. This force disrupts the detailed balance, generates stationary probability currents, and transforms the system into a purely nonequilibrium one.

Thus, microscopically, expression (4) is a regular Langevin equation (which simplifies the simulation), but macroscopically (after Zubarev selection) it turns into a controlled equation with an effective history potential and a modified information drift inducing non-Markovian behavior and a violation of detailed balance.

The duality of the Langevin and Doob concepts is formulated in this article as a conceptual interdisciplinary generalization.

**2.2. Extended Zubarev quasi-equilibrium ensemble**

The use of boundary functionals as thermodynamic variables according to Zubarev's method is realized through the modification of the quasi-equilibrium distribution operator by introducing additional local integrals of motion.

According to the principle of maximum information entropy, when fixing the average value of the functional, the modified statistical operator (the probability density in the trajectory space) takes the form:

$$\rho_q = Q^{-1}(\lambda_a, \tau)\exp\left(-\mathcal{H}_0 - \lambda_a \int_0^\tau \Theta\big(f(\mathbf{x}(t)) - a\big)dt\right),$$

where $\mathcal{H}_0$ is the standard Onsager-Mahlup action for the unperturbed trajectory, and $\lambda_a$ is the conjugate thermodynamic confining field (the intensive history parameter).

The classical nonequilibrium statistical operator (NSO) of Zubarev is based on the search for the extremum of the entropy functional S. $S = -\mathbf{Tr}(\rho \ln \rho)$.

When adding a boundary functional $\hat{F}_\Gamma$ (defined on the process trajectory over time τ as a new macroscopic variable), an additional condition of a fixed mean value $\langle \hat{F}_\Gamma \rangle = F_\Gamma$ is imposed on the distribution function ρ.

The variation of the Lagrange functional leads to a quasi-equilibrium distribution:

$$\rho_q = Q^{-1}\exp\left(-\sum_m F_m \hat{P}_m - \lambda_\Gamma \hat{F}_\Gamma\right), \tag{5}$$

where $\hat{P}_m$ are standard operators (energy, number of particles), $F_m$ thermodynamic forces (inverse temperature β, chemical potential), $\hat{F}_\Gamma$ is the boundary functional operator, $\lambda_\Gamma$ is the conjugate thermodynamic variable (intensive parameter), $Q = \mathbf{Tr}\exp\left(-\sum_m F_m \hat{P}_m - \lambda_\Gamma \hat{F}_\Gamma\right)$ is the generalized partition function.

The conjugate variable $\lambda_\Gamma$ is sought through the standard thermodynamic formalism of the generalized free potential $\Psi = \ln Q$. 1. From the equation of state. The value $\lambda_\Gamma$ is found as an implicit function of the macroscopic parameters by differentiating the potential: $\langle F_\Gamma \rangle = -\partial \ln Q / \partial \lambda_\Gamma = -\partial \Psi / \partial \lambda_\Gamma$. 2. Through fluctuation-dissipation relations: the second derivative determines the susceptibility of the system to a given functional (its variance in quasi-equilibrium): $\langle (\Delta F_\Gamma)^2 \rangle = \partial^2 \Psi / \partial \lambda_\Gamma^2$.

Physical meaning $\lambda_\Gamma$. This quantity acts as a "retaining field" or "chemical potential of trajectories." It shows how much the free energy (or entropy) of a system changes when topological or temporal constraints on the trajectory change.

Which functionalities are best to use and what benefits do they provide? The choice of functionalities depends on the physics of the system being studied. Three classes are the most promising:

| Functional type | Physical meaning | What does this mean for thermodynamics? |
|---|---|---|
| Maximum/Minimum trajectory $M_\tau = \max_{0 \le t \le \tau} X_t$ | Extreme fluctuations of a system over a fixed time. | Describes the thermodynamics of rare events, first-order phase transitions and destruction processes (mechanical or dielectric). |
| Time of stay in the region $\Gamma_A = \int_0^\tau \mathbb{I}(X_t \in A)dt$ | The net time spent by the system in a metastable or excited state A. | Allows us to introduce "metastability thermodynamics" and strictly separate the contributions of different kinetic phases to the total entropy. |

| Integral functionals $I = \int_0^\tau f(X_t)dt$ | Accumulated work, dissipated heat or entropy along the trajectory. | Provides a direct link to fluctuation theorems (Jarzynski, Crooks) Refs. [4, 5] within the framework of Zubarev's formalism, linking microscopic dynamics with macrovariables. |
|---|---|---|

What does this provide globally? A) Resolving hidden parameters: standard thermodynamics averages out rapid fluctuations. Boundary functionals "freeze" the process history into a statistical operator. B) A rigorous description of microscopic chaos. Transport equations are not for average fields, but for probability distributions of rare but critically important fluctuations.

### 2.3 Escher shift of measures

Let's consider what the Escher Transform/Escher Shift is. The Escher Shift (historically derived from actuarial mathematics [1932] Ref. [17] and rigorously developed in the theory of random processes) is a method of changing the probability measure of a process in order to make rare events typical.

Let us assume that we have an initial trajectory measure $\mathbb{P}_0$ of a random process X(t) on an interval $[0,\tau]$, and we are studying the trajectory functional $\hat{F}_\Gamma$. The Escher shift constructs a new (modified) measure $\mathbb{P}_\lambda$ according to the following rule:

$$d\mathbb{P}_\lambda[x(t)] = \frac{e^{-\lambda\,\hat{\mathrm{F}}_\Gamma[x(t)]}}{\mathbb{E}_{\mathbb{P}_0}\left[e^{-\lambda\,\hat{\mathrm{F}}_\Gamma}\right]} d\mathbb{P}_0[x(t)]. \tag{6}$$

What is the connection with Zubarev's thermodynamics and the physical meaning: 1. The denominator in (6) is precisely the generalized Zubarev partition function (or normalization integral $Q(\lambda,\tau) = \mathbb{E}_{\mathbb{P}_0}[e^{-\lambda\hat{F}_\Gamma}]$). 2. The exponential weight is the thermodynamic Boltzmann-Gibbs factor for trajectories. 3. Physical meaning: if in the original system $\mathbb{P}_0$ trajectories with short sojourn times or fast records were vanishingly rare, then the Escher measure shift mathematically "distorts" reality (by introducing a parameter $\lambda$ such that in the new ensemble $\mathbb{P}_\lambda$ these rare trajectories become the most probable). This "freezes" the temporal constraints of history into the statistical weight of the distribution operator. The parameters λ act as information glue, replacing integro-differential memory tails with instantaneous contour forces. This formalism is in strict interdisciplinary agreement with the mathematical theory of large deviations of Donsker–Varadan (Gärtner–Ellis theorem) Refs. [18, 24] and the concept of trajectory thermodynamics of s-ensembles Ref. [19].

For Markov processes, the Escher shift of measures in the trajectory space is the exact mathematical operation that is realized at the level of Fokker-Planck differential equations through the Doob's transform (h-transform) (10) Ref. [23]. Zubarev's method, by maximizing the information entropy for a fixed mean functional, actually performs the Escher shift "blindly," using physical intuition.

### 2.4 Equivalence of approaches

Let us carry out a mathematical justification of the coincidence of Zubarev's method with the Donsker-Varadan theory of large deviations Refs. [18, 24] and the Escher shift of measures in the theory of random processes.

Zubarev's NSO method using boundary functionals allows one to incorporate the system's prehistory into a statistical ensemble, thereby "freezing" information about rapid fluctuations into a thermodynamic parameter. This approach moves from the description of mean fields to equations for generating functionals that describe the full statistics of rare large deviations, not just the Gaussian distribution.

In stochastic thermodynamics, "freezing" a boundary functional means transforming the dynamic history of a process (dependent on the trajectory) into a fixed macroscopic state parameter. The method introduces a "holding field" that captures the history of fluctuations within the thermodynamic potential, allowing for non-Markovian effects (system memory) to be taken into account without complicating the microscopic equations. The term "freezing" is used here not to imply staticity in time, but to refer to the fixation of trajectory information.

In Zubarev's classical method, macroparameters $F_m(t)$ and their associated forces $\lambda_m(t)$ are time-dependent. The quasi-equilibrium operator adapts to the current time t. When we introduce a boundary functional (for example, the maximum of a trajectory $M_\tau$ over time τ), two things happen: A. The parameter fixes history. The value of the functional $M_\tau$ or the time of first reaching $\tau_{FPT}$ it "freezes" everything that happened to the system over the interval $[0,\tau]$. For a given interval, this is a fixed "history number." B. The macrovariable evolves. If we shift the observation window or consider the current time t as the upper boundary of the functional, the associated force $\lambda_\Gamma(t)$ becomes time-dependent. In nonequilibrium thermodynamics, such parameters are called quasi-integrals of motion. Their average values change slowly compared to rapid microscopic fluctuations, which allows us to apply the thermodynamic formalism at each moment t, despite the apparent non-stationarity.

One might wonder how the theory of random processes and the theory of large deviations relate. In the theory of random processes, the theory of large deviations was rigorously formulated mathematically.

The equivalence of the approaches is based on the fact that the physical “large deviation formalism in nonequilibrium thermodynamics” and the mathematical “large deviation theory (LDT Ref. [18]) in probability theory” are one and the same thing, written in different notations.

Let us show how the mathematical objects of the theory of random processes are transformed into Zubarev’s thermodynamic quantities.

1. Mathematicians formulate the problem this way: if we have a Markov process X(t), then the fraction of time it spends in the region $\Omega_a$ obeys the ergodic theorem. But over finite (albeit large) times $\tau$, the average value $g_a = \frac{1}{\tau}\int_0^\tau \Theta(X(t)-a)dt$ fluctuates.

The theory of random processes states that the probability of seeing an atypical value $g_a$ decays exponentially: $\mathbb{P}\left(\frac{1}{\tau}\int_0^\tau \Theta(X(t)-a)dt \approx g_a\right) \sim e^{-\tau I(g_a)}, \quad \tau \to \infty$.

The mathematical function $I(g_a)$ is called the rate function.

2. Translation Dictionary: Theory of Processes $\Leftrightarrow$ Zubarev's Thermodynamics. When physicists construct nonequilibrium thermodynamics according to Zubarev, they use the principle of maximum entropy, which automatically reproduces the structure of the Donsker-Varadan theory:

| Object in the theory of random processes (Donsker-Varadan) | Thermodynamic object (Zubarev/Stochastic thermodynamics) | Physical meaning |
|---|---|---|
| Esscher shift/Girsanov tilt | Extended quasi-equilibrium ensemble | Shifting the probability measure of trajectories using a weighting factor $e^{-\lambda_a \hat{\Gamma}_a}$ |
| Logarithmic cumulant generator: $\Lambda(\lambda_a) = \lim_{\tau\to\infty} \frac{1}{\tau} \ln \mathbb{E}[e^{-\lambda_a \hat{\Gamma}_a}]$ | Generalized free potential (analog of pressure or free energy) | Thermodynamic potential generating moments (cumulants) of the distribution of the history functional |
| The function of evasive speeds $I(g_a)$ | Nonequilibrium trajectory entropy (Legendre function) | System rigidity. This reflects the "information cost" (entropy production) of maintaining the system in a rarefied state |
| Slope parameter of the measure $\lambda_a$ | Conjugate thermodynamic force | An intensive parameter that plays the role of a holding field for the history of fluctuations |

3. The link between the theory of processes and Zubarev's thermodynamics is the Gartner–Ellis theorem Ref. [18]. From the side of the theory of random processes, the highest eigenvalue of the Feynman–Katz operator (9) $\Lambda_0(\lambda_a)$ is found and, according to the Gartner–Ellis theorem, it is stated that the deviation function (the probability of a rare event) is equal to the Legendre transform of this spectrum:

$$I(g_a) = \sup_{\lambda_a}[-\lambda_a g_a - \Lambda_0(\lambda_a)].$$

From the thermodynamic side, the physicist maximizes entropy according to Zubarev, introduces the Lagrange multiplier $\lambda_a$, constructs free energy and, through the thermodynamic Legendre transformation, obtains the same potential $I(g_a)$ as a function of the order parameter $g_a$.

The illusion of distinction arises because standard textbooks on the theory of random processes often omit large deviation theory Ref. [18], restricting themselves to the Gaussian approximation (the central limit theorem) and the Fokker-Planck equations for average values. However, in the modern mathematical physics of stochastic systems, these approaches have merged.

Thus, Zubarev's maximum entropy method for trajectories is simply a physical way of encoding the rigorous mathematical theory of large deviations Ref. [18] for additive functionals of Markov processes.

The violation of the Gaussian approximation when moving from small to large deviations within the sojourn time functional is a key physical node dividing the standard quasi-equilibrium thermodynamics of Onsager and the strict nonequilibrium thermodynamics of Zubarev.

Below is a detailed mathematical and physical analysis of this transition, formulated as a kinetic transition from Gaussian fluctuations to large deviations.

The central limit theorem (CLT) for random processes guarantees that for finite but sufficiently small observation times τ, fluctuations in the specific residence time $g_a$ around the mean value $g_a^{(0)}$ obey a normal law. However, as the system moves away from the local equilibrium point, nonlinear effects kick in, radically altering the symmetry and topology of the distribution.

In the small-deviation regime (the Gaussian region), the Feynman-Katz spectral equation (9) is weakly perturbed in the vicinity of $\lambda_a \to 0$. In terms of Zubarev's thermodynamics, this corresponds to a linear nonequilibrium response. The trajectory entropy (the Donsker-Varadan function of multiple deviations) is strictly parabolic: $I(g_a) \approx (g_a - g_a^{(0)})^2 / 2\chi_a$. Here, the probability distribution is symmetric with respect to the sign of the deviation $\Delta g_a = g_a - g_a^{(0)}$. In this region, the classical Onsager fluctuation-dissipation relation is fully satisfied, and the thermodynamic potential $\mathcal{F}(g_a)$ behaves like a conventional quadratic free energy potential.

Let us now discuss the mechanism of the Gaussian approximation violation (physical asymmetry). The dwell time above the level is a bounded functional: by definition, $g_a \in [0,1]$. The Gaussian distribution has infinite "tails," which directly contradicts the physical boundaries of the functional as τ increases. The violation of the Gaussian approximation occurs asymmetrically for two opposing regimes:

A. Right tail of the distribution ($g_a \to 1$, strong reinforcement). When the conjugate field is $\lambda_a \to -\infty$, the system is forced to spend almost all its time above the level *a*. In terms of the Fokker-Planck equation, this is equivalent to the generation of a powerful drift, which "presses" the particle to the upper outer boundary of the system x=L. A mathematical shift occurs: the principal eigenvalue $\Lambda_0(\lambda_a)$ depends asymptotically linearly on the field: $\Lambda_0 \propto -\lambda_a$.

A consequence for entropy: the Legendre transform hits the kinetic limit at this point. The velocity function $I(g_a)$ slopes steeply upward, becoming a vertical wall at $g_a = 1$. The probability of a rare event here decays exponentially faster than predicted by the Gaussian law.

B. The left tail of the distribution ( $g_a \to 0$ , strong penalty). This regime is of greatest interest for nonequilibrium thermodynamics and is directly related to the concept of "freezing" history. When $\lambda_a \to +\infty$, being in the region x>a becomes energetically "lethal" for trajectories (infinite probability sink).

The effect of fluctuation quantization is observed: the system is pushed out of region II. When $\lambda_a \to +\infty$, the Feynman-Katz spectral equation (9) for region I x<a turns into the problem of finding the ground state of a particle in a well with a perfectly absorbing wall at the point x=a.

Reaching a plateau (confinement): the highest eigenvalue $\Lambda_0$ ceases to depend on $\lambda$ and saturates at the fundamental quantum diffusion constant: $\Lambda_0(\lambda_a \to \infty) \to -\pi^2 D / 4a^2$. This leads to the fact that Zubarev's trajectory entropy (nonequilibrium potential) $I\left(g_a\right)$ reaches a fixed plateau at: $I(g_a \to 0) \to \pi^2 D / 4a^2$.

A thermodynamic consequence of the violation of the Gaussian approximation. Potential $I\left(g_a\right)$ saturation at $g_a \to 0$ signifies complete nonlinearity of the system. The second derivative of the deviation rate function (which in the Gaussian domain defined the mutual susceptibility $\chi_a$ ) tends to infinity: $d^2 I(g_a) / dg_a^2 \to \infty$ at $g_a \to 0$.

From the perspective of Zubarev's thermodynamics, this signifies a critical slowing of fluctuations (a kinetic phase transition) in trajectory space. The traditional fluctuation-dissipation theorem is completely destroyed here: the system becomes infinitely resistant to attempts to further reduce the residence time, since the topology of the phase space has already changed—the region x>a is completely "cut off" by effective transport drift.

Transformation of entropy production. In standard stochastic thermodynamics Ref. [1], the entropy production in a medium along a Langevin trajectory over time $\tau$ is determined by integrating the velocity fields: $\Delta s_m = \int_0^\tau \frac{\mu(x) - \partial_x D}{D} \circ \dot{x} dt$.

When moving to a distribution operator with a boundary functional and the conjugate field $\lambda_\Gamma$, the total rate of entropy production of the system $\dot{S}_{tot}$ is split into three components.

New structure of entropy production: $\dot{S}_{tot}(t) = \dot{S}_{sys}(t) + \dot{S}_{medium}(t) + \dot{S}_{history}(\lambda_\Gamma)$.

Let us write each part in terms of a modified distribution $\tilde{\rho}(x,t)$ and a normalized operator: $\rho_q(x,t) = \tilde{\rho} / Q$. 1. Change in the entropy of the system $\dot{S}_{sys}$: $\dot{S}_{sys} = -\frac{d}{dt} \int \rho_q \ln \rho_q \, dx$. 2. The classical entropy flow into the environment $\dot{S}_{medium}$ is determined through modified probability currents $J_\lambda(x,t) = \mu(x)\rho_q - D\partial\rho_q / \partial x$: $\dot{S}_{medium} = \int (\mu(x)/D) J_\lambda(x,t) dx$. 3. The trajectory contribution of history $\dot{S}_{history}$ is a fundamentally new term arising 0from fixing the boundary functional by the conjugate field: $\dot{S}_{history}(\lambda_\Gamma) = \lambda_\Gamma \cdot d\langle \hat{F}_\Gamma \rangle_q / dt - d \ln Q / dt$.

The physical consequences of the structural change include a violation of the fluctuation-dissipation theorem (FDT). The term $\dot{S}_{history}$ modifies the standard Onsager reciprocity relations. The presence of the conjugate field acts as an effective "information pump."

Generalized fluctuation theorems. The modified Fokker-Planck operator guarantees that the integral fluctuation theorem $\langle e^{-\Delta s_{tot}} \rangle = 1$ continues to hold for complete entropy production, but now it explicitly takes into account the weighting coefficients of rare fluctuations, determined by the field $\lambda_\Gamma$. The introduction of a boundary functional effectively shifts the ensemble's "equilibrium point" to the region of extreme trajectories.

Microscopic justification: a probabilistic approach. In the theory of random processes, the distribution of additive trajectory functionals is studied without explicitly introducing thermodynamic ensembles. We will show that the mathematical apparatus of the theory of Markov processes leads to identical equations of state.

Generalized partition function as a functional on a measure of trajectories. Let be $\Omega = C([0,\tau];\mathbb{R})$ the space of continuous trajectories x(t) on the interval $[0,\tau]$, and let be $\mathbb{P}_0$ the probability measure (distribution) of the trajectories of the original Langevin process. In the theory of random processes, the mathematical expectation of any functional $\hat{\Gamma}_a[x(t)]$ is defined as the integral over this measure:

$$Q(\lambda_a,\tau) = \mathbb{E}_{\mathbb{P}_0}\left[e^{-\lambda_a\hat{\Gamma}_a}\right] = \int_\Omega e^{-\lambda_a\int_0^\tau \Theta(x(t)-a)dt}\, d\mathbb{P}_0[x(t)].$$

From the point of view of probability theory, $Q(\lambda_a,\tau)$ it is the generating function of the moments of the distribution of a random variable $\hat{\Gamma}_a$.

According to the fundamental theorem of Girsanov [21] (in the physical context, the Onsager-Mahlup representation), the density of the unperturbed measure $d\mathbb{P}_0$ is expressed through the action $\mathcal{H}_0[x(t)]$:

$$d\mathbb{P}_0[x(t)] = \mathcal{D}[x(t)]\exp\left(-\int_0^\tau \frac{(\dot{x}-\mu(x))^2}{4D}dt\right).$$

Substituting this into the path integral, we obtain:

$$Q(\lambda_a,\tau) = \int_\Omega \mathcal{D}[x(t)]\exp\left(-\mathcal{H}_0[x(t)] - \lambda_a\int_0^\tau \Theta(x(t)-a)dt\right).$$

This expression is mathematically identical to the definition of the generalized Zubarev partition function in the extended quasi-equilibrium ensemble (Section 2.5).

Katz's theorem and Markov semigroups. In the theory of random processes, Katz's theorem (related to the Feynman-Katz formula Ref. [22]) is used to calculate such functional averages. If the original process X(t) is Markov with generator $\hat{L}$, then the function:

$$u(x_0,\tau) = \mathbb{E}_{\mathbb{P}_0}\left[e^{-\lambda_a\int_0^\tau \Theta(X(t)-a)dt}\,\middle|\,X(0)=x_0\right]$$

satisfies the inverse Kolmogorov equation with the particle "death" potential:

$$\frac{\partial u}{\partial \tau} = L^\dagger u - \lambda_a\Theta(x_0-a)u, \quad u(x_0,0)=1.$$

The partition function is obtained by averaging over the initial distribution $\rho_0(x_0)$:

$$Q(\lambda_a,\tau) = \int_{-\infty}^{+\infty} u(x_0,\tau)\rho_0(x_0)dx_0 = \int_{-\infty}^{+\infty} e^{L_a^\dagger\tau}\cdot 1\cdot\rho_0(x_0)dx_0.$$

By passing to the conjugate operator $\hat{L}_a$, we immediately return to the direct Fokker-Planck equation for the modified density $\tilde{\rho}(x,\tau) = e^{\hat{L}_a\tau}\rho_0(x)$.

Equivalence of equations of state. In probability theory, the mean value of a random variable $\hat{\Gamma}_a$ with a modified measure (the so-called Girsanov tilt Ref. [25] or Escher transform Ref. [26]) is sought through the logarithmic derivative of the generating function:

$$\left\langle\hat{\Gamma}_a\right\rangle_\lambda \equiv \frac{\mathbb{E}_{\mathbb{P}_0}\left[\hat{\Gamma}_a e^{-\lambda_a\hat{\Gamma}_a}\right]}{\mathbb{E}_{\mathbb{P}_0}\left[e^{-\lambda_a\hat{\Gamma}_a}\right]} = -\frac{\partial\ln Q(\lambda_a,\tau)}{\partial\lambda_a}.$$

In the limit $\tau\to\infty$, using the ergodic theorem for Markov semigroups, we find that the spectrum of the operator $\hat{L}_a$ dominates the evolution. Since the highest eigenvalue $\Lambda_0(\lambda_a)$ is isolated, the measure is concentrated on the invariant state of the operator $\hat{L}_a$. Then the specific mean sojourn time $g_a$ is strictly equal to $g_a = \lim_{\tau\to\infty}\langle\hat{\Gamma}_a\rangle_\lambda / \tau = -d\Lambda_0(\lambda_a)/d\lambda_a$.

Conclusion of equivalence. The probabilistic approach (based on measure theory on the space of trajectories and Markov semigroups) and Zubarev's phenomenological method (based on maximizing information entropy under macroscopic constraints) yield an identical free energy structure $\Lambda_0(\lambda_a)$ and identical equations of state.

Zubarev's thermodynamics acts as a macroscopic projection of the rigorous theory of large deviations of random processes, where the conjugate parameter changes the topology of the probability measure, transferring rare fluctuations of the original process into the category of typical (most probable) events of a new deformed ensemble.

**2.5. Generalized partition function**

The introduction of the functional of the process X(t) being time above a given level *a* on an interval $[0,\tau]$ allows us to construct the exact structure of the generalized partition function Q.

The formal definition of the time functional of staying above level *a* for time τ is given as the integral functional in (2). We define the generalized partition function $Q(\lambda_a,\lambda_{\mathbf{FPT}},\lambda_M)$ in terms of the path integral. The generalized partition function is a moment generator (or characteristic function) for a given trajectory functional. It is calculated using the functional integral (path integral) with the Onsager-Mahlup weight or through the trace of the modified operator: $Q(\lambda_a,\tau)=\mathbf{Tr}\left(\rho_0 e^{-\lambda_a\hat{\Gamma}_a}\right)=\int\mathcal{D}[x(t)]P[x(t)]\exp\left(-\lambda_a\int_0^\tau\Theta(x(t)-a)dt\right)$,

where $\lambda_a$ is the conjugate thermodynamic field (an intensive parameter that penalizes or rewards staying above the level), $P\left[x(t)\right]$ is the base probability density function of the trajectory in the absence of the conjugate field, and $\rho_0$ is the initial distribution of the system in phase space at t=0.

Calculation via a modified operator (Katz method). In explicit form for a diffusion process, $Q(\lambda_a,\tau)$ it is expressed by integrating the modified probability density $\tilde{\rho}(x,\tau)$ over the entire state space. According to the Feynman-Katz formula (9) Ref. [22], the partition function takes the form: $Q(\lambda_a,\tau)=\int_{-\infty}^{+\infty}\tilde{\rho}(x,\tau)dx$.

In this case, the non-normalized density $\tilde{\rho}(x,\tau)$ itself satisfies the modified Fokker-Planck equation (3), (9) with coordinate-dependent stock.

Representation in the limit of long times $\tau\to\infty$. For stationary processes in the limit of long-term observation, the asymptotic behavior of the partition function is determined by the largest (principal) eigenvalue $\Lambda_0(\lambda_a)$ of the modified operator $\hat{L}_a=\hat{L}-\lambda_a\Theta(x-a)$: $Q(\lambda_a,\tau)\propto e^{\Lambda_0(\lambda_a)\tau}, \quad \tau\to\infty$.

From the point of view of nonequilibrium thermodynamics, the quantity $\Lambda_0(\lambda_a)$ plays the role of a generalized free potential (pressure) in the space of trajectories.

The generalized partition function for the time spent above level *a* is written in terms of the functional average over the trajectories or in terms of the solution of the modified Fokker-Planck equation: $Q(\lambda_a,\tau)=<\exp\left(-\lambda_a\int_0^\tau\Theta(X(t)-a)dt\right)>=\int_{-\infty}^{+\infty}e^{\hat{L}_a\tau}\rho_0(x)dx$.

Joint generating function and extended operator. We define the generalized Zubarev partition function for three functionals as the joint functional average over the trajectories of a process X(t) starting from a point $x_0$ with an initial record $m_0\geq x_0$:

$$Q(x_0,m_0,\tau;\lambda_a,\lambda_{\mathbf{FPT}},\lambda_M)=\mathbb{E}_{x_0,m_0}\left[\exp\left(-\lambda_a\int_0^\tau\Theta(X(t)-a)dt-\lambda_{\mathbf{FPT}}\tau-\lambda_M M_\tau\right)\right]. \tag{7}$$

To reduce this problem to a deterministic differential equation, we must fix the current record $M_t = m$. Then the particle's trajectory X(t) is always bounded above by the value of the current record: $X(t) \le m$. Once the particle reaches the point x=m, any further movement to the right instantly increases the value of the coordinate m.

Let's define an auxiliary function $u(x,m,\tau)$, where x is the current coordinate and m is the current maximum. According to the theory of Markov processes with reflection/absorption at a moving boundary, the inverse Kolmogorov equation takes the form:

$$\frac{\partial u(x,m,\tau)}{\partial \tau} = \hat{L}^{\dagger} u(x,m,\tau) - \left[\lambda_a \Theta(x-a) + \lambda_{\mathbf{FPT}}\right] u(x,m,\tau)\,, \quad (8)$$

where $\hat{L}^{\dagger} = D\partial^2 / \partial x^2 + \mu(x)\partial / \partial x$ is unperturbed inverse operator.

Including the absolute maximum of a trajectory in the base of nonequilibrium thermodynamic variables elevates the description to the level of nonlocal functionals, requiring an expansion of the phase space. Using the Stroock-Varadhan boundary condition (relation (11), Section 3) allows us to relate peak fluctuations to the thermodynamic potential, while the resulting non-Markovian drift effectively describes the system's adaptation to historical records.

## 3. Dynamic equations in extended phase space and connection with the theory of random processes

Zubarev's classical NSO formalism traditionally operates with time-local macrovariables, the evolution of which is determined by the Liouville or Fokker-Planck equations. Incorporating history functionals into the thermodynamic basis—such as the FPT, the time spent above the level, and the absolute extremum of the trajectory—requires a fundamental restructuring of the mathematical apparatus for describing evolution.

The central task here is the transition from continuous integration over the space of trajectories (the Langevin–Onsager–Mahlup representation) to deterministic partial differential equations for probability densities (the Liouville–Fokker–Planck representation).

This transition relies on the duality between the physical principle of maximum information entropy and the mathematical apparatus of Markov semigroups. To rigorously relate the conjugate thermodynamic forces $\lambda_a, \lambda_{\mathbf{FPT}}, \lambda_M$ to the measurable order parameters of the system, we develop a series of equations describing: a) the evolution of the unnormalized trajectory measure under the pressure of local and global penalties (modified Feynman-Katz equation); b) kinetic constraints at the phase space boundaries generated by nonlocal record statistics (Stroock-Varadhan boundary conditions); c) the resulting macroscopic transport of mass and heat in the modified physical space (generation of non-Markovian drift via the Doob h-transformation).

Below is a closed system of these equations, which describes in detail the kinetics of a three-parameter nonequilibrium ensemble in the extended phase space (x, m), where x is the current coordinate and m is the current historical maximum of the process.

### 3.1 Generalized partition function and the Feynman-Katz equation.

The generalized Zubarev partition function $Q(\lambda_a,\tau) = \langle e^{-\lambda_a \hat{\Gamma}_a} \rangle$ is expressed through the trace of a modified evolution operator. The unnormalized probability density $\tilde{\rho}(\mathbf{x},\tau)$ satisfies the direct Feynman-Katz spectral equation:

$$\frac{\partial \tilde{\rho}(\mathbf{x},\tau)}{\partial \tau} = \hat{L}\tilde{\rho}(\mathbf{x},\tau) - \lambda_a \Theta\big(f(\mathbf{x}) - a\big)\tilde{\rho}(\mathbf{x},\tau) \quad (9)$$

where $\hat{L} = D\nabla^2 - \nabla \cdot (\mathbf{F}\cdot)/\gamma$ is the standard Fokker-Planck operator. The thermodynamic field acts here as a local absorption/fictitious sink potential.

The simultaneous introduction of two boundary functionals—the time of first arrival $\tau_{\mathbf{FPT}}$ and the time spent above the level $\hat{\Gamma}_a$—generates a spatiotemporal separation of trajectory fluctuations, which cannot be described by either parameter separately. Within the framework of Zubarev's method, this corresponds to the introduction of two conjugate fields $\lambda_{\mathbf{FPT}}$ and $\lambda_a$, which form a two-parameter nonequilibrium thermodynamics system.

Cross-correlations and Onsager symmetry breaking. The main mathematical effect is the emergence of mixed susceptibility $\chi_{\mathbf{cross}}$. In the Gaussian approximation (linear response), the susceptibility matrix becomes off-diagonal:

$$\begin{pmatrix} \Delta\tau_{\mathbf{FPT}} \\ \Delta g_a \end{pmatrix} = -\begin{pmatrix} \chi_{\mathbf{FPT}} & \chi_{\mathbf{cross}} \\ \chi_{\mathbf{cross}} & \chi_a \end{pmatrix}\begin{pmatrix} \lambda_{\mathbf{FPT}} \\ \lambda_a \end{pmatrix},$$

where the cross-term is determined by the integral of the mutual correlation function:

$$\chi_{\mathbf{cross}} \propto \int_0^\infty \langle \Delta\tau_{\mathbf{FPT}}(0)\Delta\Theta(X(t)-a)\rangle_0\, dt\,.$$

Physical meaning: This term describes how the system's early history (how quickly it reached level $a$) rigidly determines its subsequent thermodynamic behavior (how long it will spend above it). Under highly nonequilibrium conditions, the sign $\chi_{\mathbf{cross}}$ allows for a precise diagnosis of the arrow of time and the degree of disruption of the detailed balance.

Two-parameter effective transport (Doob`s transform). The introduction of two fields radically changes the structure of the modified Fokker-Planck equation. According to the Feynman-Katz formula, the joint generation of the distribution is described by the operator:

$$\hat{L}_{\mathbf{double}} = \hat{L} - \lambda_a\Theta(x-a) - \lambda_{\mathbf{FPT}}\,.$$

In this case, the field $\lambda_{\mathbf{FPT}}$ acts as a global spectral shift (changing the “lifetime” of trajectories), and $\lambda_a$ as a local spatial sink.

Applying the Doob`s transformation (h-transform, Section 3.2, with drift (10)) to find the effective drift $v_{eff}(x)$ leads to interference of forces:

$$v_{eff}(x) = F(x)/\gamma + 2D\partial \ln\phi_0(x;\lambda_a,\lambda_{\mathbf{FPT}})/\partial x\,. \tag{10}$$

Two-dimensional nonequilibrium potential and phase transitions. In large deviation theory (LDT), the joint deviation rate function (two-dimensional trajectory entropy) $I(g_a,\tau_{\mathbf{FPT}})$ is obtained by double Legendre transform of the principal eigenvalue $\Lambda_0(\lambda_a,\lambda_{\mathbf{FPT}})$:

$$I(g_a,\tau_{\mathbf{FPT}}) = \lambda_a g_a + \lambda_{\mathbf{FPT}}\tau_{\mathbf{FPT}} + \Lambda_0(\lambda_a,\lambda_{\mathbf{FPT}})\,.$$

This entropy landscape in the nonlinear regime $\lambda \to \infty$ reveals geometric features (folds, swallowtail singularities), indicating dynamic (trajectory) first-order phase transitions.

The system discretely switches between two macroscopic modes: "slow diffusion with long retention" and "fast ballistic jump with instantaneous release." Single-parameter models (with only FPT or only residence time) are unable to detect these transitions, as they average the second coordinate.

Adding two functionals simultaneously allows us to separate the effects of the speed of reaching a state and the duration of its retention. This yields an accurate cross-correlation matrix (Onsager test), two-parameter control of traffic flows, and allows us to detect hidden trajectory phase transitions in the two-dimensional landscape of the nonequilibrium potential $I(g_a,\tau_{\mathbf{FPT}})$..

For the mathematical description of Zubarev's three-parameter nonequilibrium thermodynamics, including additive functionals $(\hat{\Gamma}_a,\tau_{\mathbf{FPT}})$ and a non-local extremum ($M_\tau = \max_{0\le t\le\tau} X(t)$), the standard apparatus of the Feynman-Katz formula requires modification.

Since the trajectory maximum $M_\tau$ is a monotonically non-decreasing process $(M_t = \max\{M_0, \max_{0\le s\le t} X(s)\})$, the joint evolution of the probability density is described in an extended phase space, where the current value of the record acts as an additional dynamic coordinate.

Specific boundary conditions for the extremum. The influence of the conjugate field of extrema $\lambda_M$ is fully transferred to the dynamic boundary condition at the current record point x=m.

When the particle reaches the boundary x=m, the time increment (dt) and the record increment (dm) are stochastically related. Applying the Ito lemma to the phase space boundary, we obtain the Stroock-Varadhan boundary condition Ref. [27] for the field $\lambda_M$:

$$\left.\frac{\partial u(x,m,\tau)}{\partial x}\right|_{x=m} = \lambda_M u(m,m,\tau). \tag{11}$$

The physical meaning of the boundary condition. At $\lambda_M > 0$ (penalty for high records), condition (11) becomes a partially absorbing boundary (Robin`s boundary condition Ref. [28]). The field $\lambda_M$ acts as an "effective elastic wall" that dampens the statistical weight of trajectories attempting to break the record. At $\lambda_M \to \infty$, condition (11) becomes purely absorbing $(u(m,m,\tau)=0)$. This means that trajectories are strictly forbidden from exceeding the level m, and Zubarev's thermodynamic ensemble completely isolates the system below this level.

Three-dimensional landscape of a nonequilibrium potential. For a stationary process in the limit $\tau \to \infty$, spectral analysis of this operator yields the principal eigenvalue $\Lambda_0(\lambda_a, \lambda_{\mathbf{FPT}}, \lambda_M)$, which now depends on the geometric shape of the boundary.

The transition to macroscopic coordinates of history (the fraction of time $g_a$, the transit time $\tau_{\mathbf{FPT}}$ and the magnitude of the peak fluctuation shock M) is carried out through a multidimensional equation of state:

$$g_a = -\frac{\partial \Lambda_0}{\partial \lambda_a}, \quad \tau_{\mathbf{FPT}} = -\frac{\partial \Lambda_0}{\partial \lambda_{\mathbf{FPT}}}, \quad M = -\frac{\partial \Lambda_0}{\partial \lambda_M}.$$

The full nonequilibrium potential (Zubarev's three-dimensional trajectory entropy) is constructed through the generalized Legendre transformation:

$$\mathcal{F}(g_a, \tau_{\mathbf{FPT}}, M) = \lambda_a g_a + \lambda_{\mathbf{FPT}} \tau_{\mathbf{FPT}} + \lambda_M M + \Lambda_0(\lambda_a, \lambda_{\mathbf{FPT}}, \lambda_M).$$

Non-Markovian induced transport. Applying the Doob`s h-transformation with drift (12) to the extended operator, the effective particle drift velocity $v_{eff}$ under the action of three conjugate Zubarev fields is written as:

$$v_{eff}(x,m) = \mu(x) + 2D\partial \ln \phi_0(x,m;\lambda_a, \lambda_{\mathbf{FPT}}, \lambda_M)/\partial x, \tag{12}$$

where $\phi_0(x,m)$ is the left-hand principal eigenfunction of the modified evolution operator, explicitly depending on the current record m. This leads to the effect of dynamic adaptation of the forces. As long as the current coordinate x is far from the historical maximum m, the gradient $\partial_x \ln \phi_0$ is small, and the particle moves under the action of standard forces and the sojourn time field. As soon as $x \to m$, exponential deceleration (or acceleration, depending on the sign of $\lambda_M$) kicks in, governed by boundary condition (11).

This is mathematical confirmation that introducing extremes into thermodynamics "freezes" history not in the form of a constant drift, but in the form of an effective non-Markovian memory that reads the current status of the system's records.

Figure 1 illustrates the behavior of expression (12) for $v_{\mathbf{eff}}(x,m)$. This graph shows how the force returning the particle depends on the proximity of the coordinate to the historical record m. Here, the first term μ=$-1\cdot x$ is the initial Langevin force (the usual linear restoring force of a harmonic oscillator), and the second term is the non-Markovian impact when approaching the record. We set $\lambda_a$=100, D=1. The x-axis range is from -2 to 3, with a step of 0.05. The formula models the force that “remembers” the particle’s previous records. The second term contains a multiplier $-2\sqrt{D\lambda_a}$ (the impact amplitude) and a decaying exponential exp(-(m-x)/0.2). The coefficient 0.2 in the denominator of the exponent was selected empirically for Mathcad. It determines the radius of the "memory". Three curves were plotted for different values of the record m: $v_{eff}(x,0.5,100)$ (m=0.5, curve 1 in Figure 1, solid green line), $v_{eff}(x,1.5,100)$ (m=1.5, curve 2 in Figure 1, dotted black line), and $v_{eff}(x,2.5,100)$ (m=2.5, curve 3 in Figure 1, dashed-dotted red line). The graph clearly shows that while x is far from m, both curves follow the linear Langevin trend. However, as soon as x approaches its record point ($x \to 0.5$ for the first curve, $x \to 1.5$ for the second, and $x \to 2.5$ for the third), the speed graph drops sharply and singularly. This is a graphical demonstration of the "frozen-in memory" of the extremum.

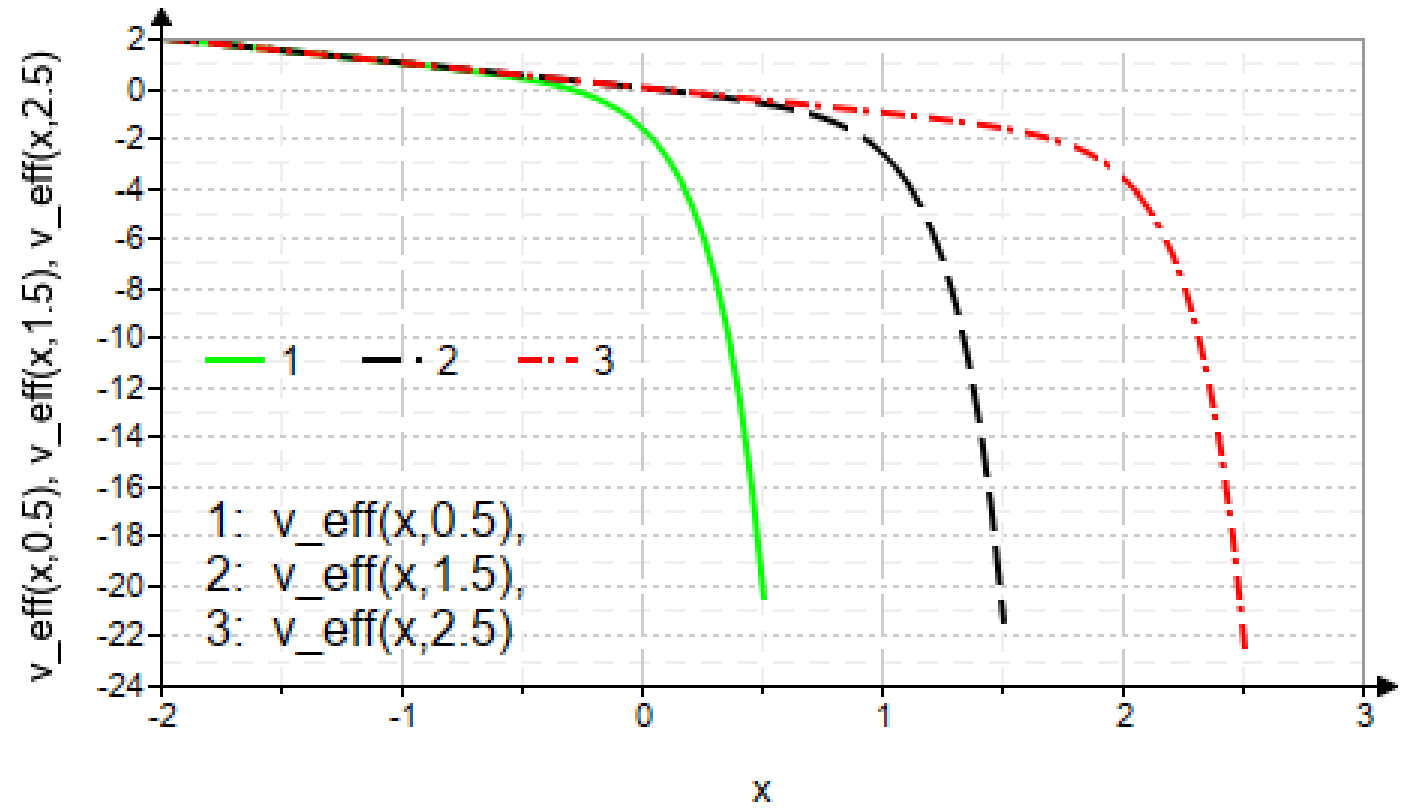

Fig.1. The induced effective non-Markovian drift velocity $v_{\mathbf{eff}}(x,m)$ as a function of the current coordinate x, computed via Doob's h-transform for three distinct historical records: m=0.5 (solid green line), m=1.5 (dashed black line), and m=2.5 (dotted red line). At a distance from the record $x \ll m$, all curves collapse onto the classical linear Langevin restoring force $\mu(x) = -x$ of a harmonic oscillator. As the particle approaches its historical maximum $x \to m$, the conjugate thermodynamic field of extremes $\lambda_M$ triggers a singular returning force (a "memory shock"), dropping to $\approx -20$. Beyond the record boundary $x > m$, the trajectories are strictly forbidden by the Stroock-Varadhan boundary condition, represented via the mathematical termination of the curves at x=m using the NaN operator in the state-space formulation. The parameters used are D=1 and $\lambda_a = 100$.

In a real system, an absorbing or reflecting wall must be activated at the point x=m, and the function $v_{eff}$ for x>m simply doesn't exist. It has been shown that the drift force is not static. If the particle is far from its record (x<<m), the force behaves normally. But as the record approaches $x \to m$, the drift graph drops sharply (demonstrating a singular velocity return shock $v_{eff} \sim -2\sqrt{D\lambda_a}$). This clearly demonstrates the "frozen memory" of the extrema.

**3.2 Thermodynamic limit $(\tau \to \infty)$ and Legendre transformation**

In the long-term observation limit, the asymptotic behavior of the partition function is determined by the highest eigenvalue $\Lambda_0(\lambda_a)$ of the modified operator $\hat{L}_a = \hat{L} - \lambda_a \Theta(f(\mathbf{x}) - a)$ : $Q(\lambda_a, \tau) \propto e^{\Lambda_0(\lambda_a)\tau}$, $\tau \to \infty$. The quantity $\Lambda_0(\lambda_a)$ plays the role of a nonequilibrium free potential.

The transition to a specific macroscopic coordinate (fraction of residence time) $g_a = \Gamma_a / \tau$ specifies the thermodynamic equation of state: $g_a(\lambda_a) = -d\Lambda_0(\lambda_a) / d\lambda_a$.

To move to an entropic description of trajectories, the Legendre transformation is used, generating a rate function $I(g_a)$ equivalent to the trajectory entropy of the system: $\mathcal{F}(g_a) \equiv I(g_a) = \lambda_a(g_a) \cdot g_a + \Lambda_0(\lambda_a(g_a))$.

Let's consider the dynamic equations of a trajectory ensemble and their connection to the theory of random processes. Zubarev's classical NSO formalism traditionally operates with macrovariables local in time, the evolution of which is determined by the microscopic Liouville or Fokker-Planck equations.

Modified Feynman-Katz equation (volume dynamics). For the auxiliary function $u(x,m,\tau)$, which defines the joint momentum generating function, the inverse Kolmogorov equation with local and global penalty potentials takes the form:

$$\frac{\partial u(x,m,\tau)}{\partial \tau} = D\frac{\partial^2 u}{\partial x^2} + \mu(x)\frac{\partial u}{\partial x} - \left[\lambda_a \Theta(x-a) + \lambda_{\mathbf{FPT}}\right] u(x,m,\tau). \tag{13}$$

Stroock-Varadhan boundary condition (record dynamics). The influence of the conjugate thermodynamic field of extrema $\lambda_M$ is transferred to the moving boundary of the current maximum x=m in the form of the nonlocal Robin`s boundary condition (11).

The Doob`s h-transformation (induced macroscopic transport). Let us have a homogeneous Markov chain $X_t$ on a state set S with transition matrix $P(x,y) = \mathbb{P}(X_{t+1} = y | X_t = x)$. We choose a strictly positive function h: S→(0, +∞), which is harmonic for our process (or for a subset of it). The harmonic property means that $\sum_{y\in S} P(x,y)h(y) = h(x)$. The transformed transition matrix $P_h(x,y)$ is given by the formula: $P_h(x,y) = P(x,y)h(y)/h(x)$. The Doob`s h-transformation adds an additional drift to the process, directed toward increasing the function h. The new stochastic differential equation is: $dX_t = \left(b(X_t) + \sigma^2(X_t)(\nabla h(X_t)/h(X_t))\right)dt + \sigma(X_t)dW_t$. Here $\nabla h(X_t)/h(X_t) = \nabla \ln h(X_t)$, acts as a "force" that nudges the process trajectories in the desired direction.

The normalized mass current $J_\lambda(x,m,t)$ in the physical coordinate space is freed from the stock terms and obeys the continuity equation with effective drift (12), as in (10), but dependent on the record history:

$$\partial_t \rho_q = -\partial_x J_\lambda, \quad J_\lambda = v_{eff}(x,m)\rho_q - D\partial\rho_q/\partial x. \tag{14}$$

In the region of weak confining fields $(\lambda \to 0)$, the nonlinear system of equations (9), (12)–(14) is linearized due to the smallness of the deviation of the Zubarev operator from the base state, where $\ln\phi_0 \approx \ln\phi_0^{(0)} = 0$. In this limit, the effective velocity $v_{eff}(x,\ m)$ reduces to the classical Langevin force $\mu(x)$, and the contribution of the record coordinate m transforms into standard fluctuation correlators. The procedure for transition to classical Onsager thermodynamics via spectral corrections to the principal eigenvalue is described in the next section.

## 4. Linear response and the fluctuation-dissipation theorem

Equations (9), (12)–(14) are linearized for weak fields, where the history functionals $g_a, \tau_{\mathbf{FPT}}, M$ transition to a regime determined by the unperturbed properties. Application of spectral perturbation theory leads to a generalized Onsager law with a symmetric susceptibility matrix $\chi_{ij}$ expressed in terms of the Green–Kubo correlation functions.

In the vicinity of the natural (unperturbed by the confining field) state of the system, when the conjugate field intensity is small ($\lambda_a \to 0$), the response of the macroscopic variable $g_a$ can be described within the framework of linear perturbation theory. In the region of weak conjugate fields (λ→0), the thermodynamic response is described by the generalized linear Onsager law. Applying spectral perturbation theory to the spectrum of the Feynman-Katz operator in the vicinity of equilibrium leads to an equation of state relating the order parameter vector to the thermodynamic forces through a symmetric matrix of mutual susceptibilities of history (Kubo's theorem):

$$\begin{aligned}\Delta g_a &= g_a - g_a^{(0)} = -\chi_{g,g}\lambda_a - \chi_{g,FPT}\lambda_{FPT} - \chi_{g,M}\lambda_M\,,\\ \Delta\tau_{FPT} &= \tau_{FPT} - \tau^{(0)}{}_{FPT} = -\chi_{FPT,g}\lambda_a - \chi_{FPT,FPT}\lambda_{FPT} - \chi_{FPT,M}\lambda_M\,,\\ \Delta M &= M_\tau - M_\tau^{(0)} = -\chi_{M,g}\lambda_a - \chi_{M,FPT}\lambda_{FPT} - \chi_{M,M}\lambda_M\,.\end{aligned} \tag{15}$$

**4.1. Decomposition of the spectrum of the Feynman-Katz operator**

Consider a modified operator $\hat{L}_a = \hat{L} - \lambda_a \hat{V}(x)$, where the indicator function $\hat{V}(x) = \Theta(x-a)$ serves as the perturbation potential. According to the non-stationary Rayleigh-Schrödinger perturbation theory, the expansion of the principal eigenvalue $\Lambda_0(\lambda_a)$ in powers of the small parameter $\lambda_a$ has the form:

$$\Lambda_0(\lambda_a) = \Lambda_0^{(0)} + \lambda_a\Lambda_0^{(1)} + \lambda_a^2\Lambda_0^{(2)}/2 + \mathcal{O}(\lambda_a^3).$$

Since for an unperturbed Fokker-Planck operator $\hat{L}$ with free or reflecting boundaries the highest eigenvalue is strictly zero $(\Lambda_0^{(0)} = 0)$ and the corresponding right eigenvector coincides with the stationary distribution density $\psi_0^{(0)}(x) = \rho_0(x)$ (while the left eigenvector is $\phi_0^{(0)}(x) = 1$), the first- and second-order corrections are written as:

1. First order correction: $\Lambda_0^{(1)} = -\int_{-\infty}^{+\infty}\phi_0^{(0)}(x)\hat{V}(x)\psi_0^{(0)}(x)dx = -\int_a^L \rho_0(x)dx = -g_a^{(0)}$, where $g_a^{(0)}$ is unperturbed (stationary) fraction of time spent by the system above the level *a*.

2. Second order correction: $\Lambda_0^{(2)} = 2\int_{-\infty}^{+\infty}\hat{V}(x)\hat{L}_{inv}^{-1}\left[\hat{V}(x) - g_a^{(0)}\right]\rho_0(x)dx \equiv \chi_a$, where $\hat{L}_{inv}^{-1}$ is generalized inverse operator (Green operator) defined on the subspace of functions with zero mean.

**4.2 Generalized susceptibility of history**

The quantity $\chi_a$ represents the generalized trajectory susceptibility of the system to the confining field. Using the integral representation of the Green's operator through autocorrelation functions, we obtain a Kubo-type identity:

$$\chi_a = 2\int_0^\infty \langle \Delta\Theta(x(t)-a)\Delta\Theta(x(0)-a)\rangle_0\, dt\,,$$

where $\Delta\Theta(x(t)-a) = \Theta(x(t)-a) - g_a^{(0)}$, and averaging $\langle \ldots \rangle_0$ is carried out along the unperturbed ensemble of trajectories. An exact analytical calculation of the sojourn time susceptibility for a segment yields the expression: $\chi_a = 2a^2(L-a)^2/3DL$.

Substituting the resulting expansion of the potential $\Lambda_0(\lambda_a) \approx -g_a^{(0)}\lambda_a + \chi_a\lambda_a^2/2$ into the macroscopic equation of state, we find the linear response law: $g_a(\lambda_a) = -d\Lambda_0/d\lambda_a = g_a^{(0)} - \chi_a\lambda_a$.

This relationship is a trajectory analogue of the fluctuation-dissipation theorem (FDT): the quasi-equilibrium deviation of the fraction of the residence time from its base value is directly proportional to the magnitude of the conjugate force $\lambda_a$, and the proportionality coefficient is the integral of the autocorrelation function of fluctuations of the same quantity in the absence of a field.

Let us turn to the Onsager susceptibility matrix $\chi_{ij}$ for the linear response section. The inclusion of cross-correlations between fundamentally different functionals—additive and boundary—is one of the main results extending the basic model Ref. [11].

**4.3. Onsager's matrix of mutual susceptibilities in the trajectory ensemble**

In the region of weak conjugate fields $\lambda_i \to 0$, the thermodynamic response of a multiparameter system is described by Onsager's generalized linear law. For a three-parameter ensemble, including the specific residence time above the level $g_a$, the boundary first-passage time $\tau_{\mathrm{FPT}}$, and the nonlocal absolute extremum of the trajectory $M_\tau$, the thermodynamic equation of state for (15) in vector-matrix form takes the form:

$$\begin{pmatrix} g_a - g_a^{(0)} \\ \tau_{\mathbf{FPT}} - \tau_{\mathbf{FPT}}^{(0)} \\ M_\tau - M_\tau^{(0)} \end{pmatrix} = -\begin{pmatrix} \chi_{g,g} & \chi_{g,\mathbf{FPT}} & \chi_{g,M} \\ \chi_{\mathbf{FPT},g} & \chi_{\mathbf{FPT},\mathbf{FPT}} & \chi_{\mathbf{FPT},M} \\ \chi_{M,g} & \chi_{M,\mathbf{FPT}} & \chi_{M,M} \end{pmatrix} \begin{pmatrix} \lambda_a \\ \lambda_{\mathbf{FPT}} \\ \lambda_M \end{pmatrix},$$

where the superscript (0) denotes the unperturbed (natural) mean values of the functionals with the history fields completely turned off.

According to the fluctuation-dissipation theorem (FDT) and the principle of microscopic reversibility (detailed balance) of the original Langevin process, the matrix of history susceptibilities is strictly symmetric: $\chi_{ij} = \chi_{ji}$.

The structure and physical meaning of matrix elements. Each matrix element is expressed through the auto- or cross-correlation functions of the fluctuations of the corresponding functionals, calculated using the unperturbed Markov measure (Green–Kubo representation):

Diagonal elements (eigen-susceptibilities). A). $\chi_{g,g}$: generalized sojourn time susceptibility, calculated earlier by us analytically for the interval [0, L]: $\chi_{g,g} = 2\int_0^\infty \langle \Delta\Theta(t)\Delta\Theta(0)\rangle_0 \, dt = 2a^2(L-a)^2 / 3DL$. B). $\chi_{\mathbf{FPT},\mathbf{FPT}}$: first-passage time variance. It determines the kinetic rigidity of the barrier and the susceptibility of the system to the global time shift $\lambda_{\mathbf{FPT}}$. C). $\chi_{M,M}$: susceptibility of record statistics. Physically, it specifies the elasticity of the trajectory ensemble with respect to peak fluctuation impacts.

Off-diagonal elements (history cross-correlations). A). $\chi_{g,\mathbf{FPT}}$: the cross-susceptibility relating the barrier-reaching rate to the retention time above it. Mathematically, it is given by the integral: $\chi_{g,\mathbf{FPT}} = \int_0^\infty \left[ \langle \tau_{\mathbf{FPT}}(0)\Theta(X(t)-a)\rangle_0 - \tau_{\mathbf{FPT}}^{(0)} g_a^{(0)} \right] dt$. Physical meaning: This term describes the interference of the prehistory. It shows how the fluctuational acceleration of the process at the threshold stage (the change $\tau_{\mathbf{FPT}}$ correlates with the subsequent metastable retention of the particle in region II. B). $\chi_{g,M}$ and $\chi_{\mathbf{FPT},M}$: the coefficients relating the nonlocal extremum to additive time scales. They show the "information cost" of generating high records. The presence of these terms proves that introducing an extremum field $\lambda_M$ through the Stroock-Varadhan boundary conditions automatically shifts not only the mean maximum but also changes the mean residence and transit times.

Significance for particle separation (Section 6). In the particle separation problem, it is this matrix that determines the initial divergence of spatial paths. If the eigencoefficients for species A and B are close $\chi_{g,g}$, the cross-terms $\chi_{g,\mathbf{FPT}}$ will differ significantly due to the different topologies of the potential wells. This allows the experimenter, by varying the field ratio $\lambda_a / \lambda_{\mathbf{FPT}}$, to maximize the difference in effective drifts and achieve ideal spatial phase separation.

Figure 2 depicts two-dimensional cross-sections (isoline maps/contour plots) of the large deviation function $I(g_a, \tau_{\mathbf{FPT}})$ at various fixed extrema $M_\tau$, and the three-dimensional trajectory entropy landscape. It is shown how, for small extrema, the isolines have an ideal elliptical (Gaussian) shape, while, as they move away from the minimum (into the nonlinear regime), they deform, encountering rigid kinetic boundaries $(g_a = 1)$ or $(\tau_{\mathbf{FPT}} \to 0)$.

To visualize nonlinear landscape deformations, Mathcad uses a 3D contour plot. Initial parameters: D=1, $g_0$=0.5, and $\chi_a$=0.08. To ensure the plot accurately reflects nonlinearity, we add higher-order terms to the Gaussian form, simulating rigid kinetic barriers (taking into account the abutment of the boundaries $g_a \to 0$ and $g_a \to 1$): $I(g,\tau) := \frac{(g-g0)^2}{2 \cdot \chi_a} + \frac{0.01}{g \cdot (1-g)} + \frac{(\tau-1)^2}{2 \cdot 0.1}$.

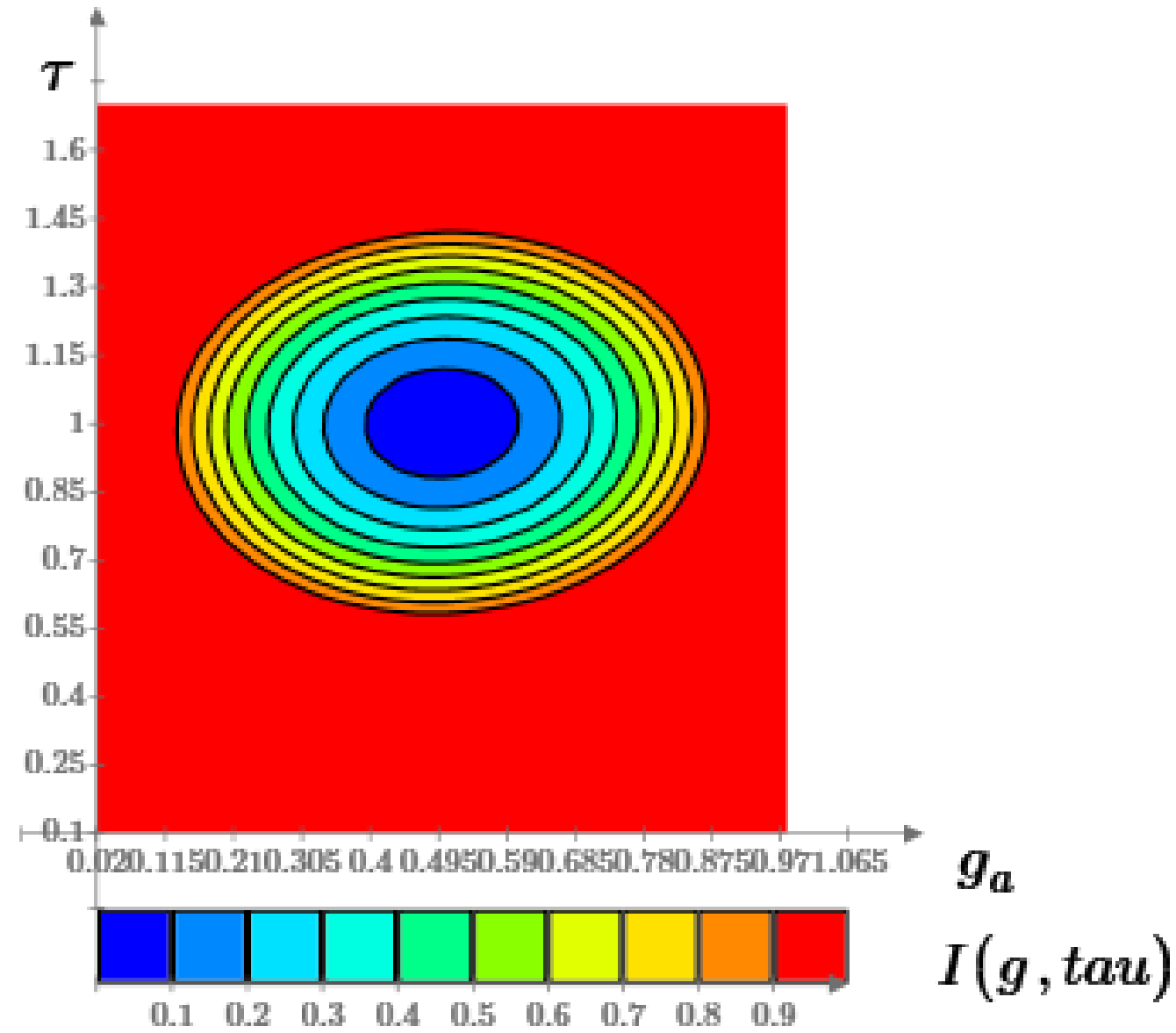

Fig. 2. The non-equilibrium thermodynamic potential $\mathcal{F}(g_a, \tau_{\mathbf{FPT}}) \equiv I(g_a, \tau_{\mathbf{FPT}})$ in the space of history descriptors (at various fixed extremes $M_\tau$), computed via the generalized Zubarev's non-equilibrium statistical operator method and Donsker-Varadhan large deviation theory. The color spectrum from blue to red is the value of the nonequilibrium trajectory entropy of Zubarev–Varadan $\mathcal{F}$. The $g_a$ center of the landscape at $(g_a = 0.5, \tau_{\mathbf{FPT}} = 1.0)$ corresponds to the local macroscopic equilibrium (minimum of the rate function). The elliptical contour lines in the central vicinity demonstrate the Gaussian regime governed by the symmetric Onsager matrix of joint susceptibilities $\chi_{ij}$. Near the boundaries $(g_a \to 0,1 \; and \; \tau_{\mathbf{FPT}} \to 0,1.75)$, the contours undergo severe non-linear deformation and contract into dense parallel lines, visually confirming the kinetic saturation effect and trajectory confinement. The global parameters are set to D=1, $g_0$=0.5, and $\chi_a$=0.08. The vertical Z-axis (represented by the color spectrum from blue to red) defines the magnitude of the non-equilibrium trajectory entropy $\mathcal{F}$ ($g_a$, $\tau_{\mathbf{FPT}}$). The blue core corresponds to the absolute minimum of the potential ($\mathcal{F}$ = 0), signifying the local macroscopic equilibrium state. The transition to the solid red plateau indicates the sharp exponential suppression of rare fluctuation paths as they approach the fundamental kinetic boundaries of the system.

The lines of constant entropy (contours) are stretched at an angle, which clearly demonstrates the existence of cross-correlations in the history of fluctuations $\chi_{g,\mathbf{FPT}}$.

## 5. Modified transport equations. Nonlinear limit of strong fields and self-similar asymptotics

The introduction of a confining field $\lambda_a$ distorts the statistical weight of trajectories. To understand what macroscopic physical forces must act on the system to maintain a nonequilibrium distribution with a fixed $g_a$, it is necessary to move to effective transport equations in physical coordinate space.

### 5.1. Doob's h-transform

Let the original Markov process obey the Fokker-Planck equation $\partial_t \tilde{\rho} = \hat{L}\tilde{\rho} - \lambda_a \Theta(x-a)\tilde{\rho}$ . This system evolves with loss of normalization. To derive a rigorous hydrodynamic equation for the normalized probability density function $\rho_q(x,t)$ in the modified ensemble, we apply the generalized Doob transform (Section 3.2). We define the normalized density function as:

$$\rho_q(x,t) = \frac{\phi_0(x)\psi_0(x)}{\int \phi_0(x)\psi_0(x)dx},$$

where $\psi_0(x)$ и $\phi_0(x)$ are the right and left principal eigenvectors of the modified operator $\hat{L}_a$ satisfying the equations:

$$\hat{L}\psi_0(x) - \lambda_a \Theta(x-a)\psi_0(x) = \Lambda_0 \psi_0(x),$$

$$\hat{L}^{\dagger}\phi_0(x) - \lambda_a \Theta(x-a)\phi_0(x) = \Lambda_0 \phi_0(x).$$

If the original dynamics is potential with a stationary Boltzmann distribution $\rho_0(x) \propto e^{-U(x)/D}$ , the left and right eigenfunctions are related by a symmetry relation (detailed balance): $\psi_0(x) = \phi_0(x)\rho_0(x)$ . Then the normalized density in the Zubarev ensemble takes the form $\rho_q(x) = \phi_0^2(x)\rho_0(x)$ .

**5.2 Continuity equation and effective drift (mass current)**

Differentiating $\rho_q(x,t)$ with respect to time, we move to a deterministic continuity equation (of mass/substance transport) without sink and source terms (14).

The effective transport velocity field (induced drift) $v_{eff}(x)$ (12) consists of the initial deterministic force and an additional thermodynamic gradient generated by the fluctuation history: $v_{eff}(x) = F(x)/\gamma + 2D\partial \ln \phi_0(x)/\partial x$ . The term $2D\partial_x \ln \phi_0(x)$ represents the thermodynamic force induced by the confining field. Since $\phi_0(x)$ monotonically decreases upon entering the "penalized" region x>a (at $\lambda_a > 0$ ), the logarithmic derivative becomes strictly negative. This means that the field $\lambda_a > 0$ generates a macroscopic return mass flux directed from the region x>a to the interface x=a, tending to forcibly reduce the residence time above the level.

**5.3 Heat transport and local dissipation**

According to the first law of stochastic thermodynamics, induced drift $v_{eff}(x)$ leads to localized heat release in the thermostat. The rate of heat transport (release) into the medium $\dot{q}(x)$ per unit volume of phase space is determined by the product of the effective thermodynamic force and the induced current:

$$\dot{q}(x) = \gamma v_{eff}(x) \cdot J_\lambda(x) = \left( F(x) + 2\gamma D \frac{\partial \ln \phi_0(x)}{\partial x} \right) J_\lambda(x).$$

By integrating this expression over the entire volume of the system, we obtain the total rate of energy dissipation required to maintain a nonequilibrium macrostate with a modified history. In a steady state, taking into account the identity $\int J_\lambda dx = 0$ for closed systems, total thermal dissipation transforms into the rate of entropy production of the environment:

$$\dot{S}_{medium} = \frac{1}{D}\int_{-\infty}^{+\infty} v_{eff}^2(x)\rho_q(x)dx.$$

In the linear response regime (according to Section 4), this integral reduces to a compact Onsager form $\dot{S}_{medium} \approx \chi_a \lambda_a^2$ , which relates the heat generation intensity to the square of the macroscopic confining history field.

### 5.4 Limit of an infinitely strong conjugate field

In the nonlinear limit of an infinitely strong conjugate field $\lambda_a \to \infty$, the system exhibits kinetic compression, where the boundary x=a becomes a perfectly reflective wall, and the residence time above the level vanishes. The information entropy of the trajectories $I(g_a)$ acts as an effective confinement potential, proving that time constraints violate Onsager's classical fluctuation-dissipation theorem.

In the nonlinear limit of a strong confining field $(\lambda \to \infty)$, the left eigenfunction of the Feynman-Katz equation (9) $\phi_0(x,m)$ acquires a self-similar form described by the universal function $\Phi((a-x)/\xi(\lambda))$, which corresponds to the generation of efficient non-Markovian Doob`s transport with divergent velocity near the boundary. The developed multiparameter approach extends the results of Ref. [11] by converting the boundary functionals into active thermodynamic variables and overcomes the limitations of the one-parameter description of the first-passage time from Refs. [12–16] by introducing history cross-correlations and identifying trajectory phase transitions.

In the nonlinear regime, with strong overshoot penalties $\lambda_a \to +\infty$, the scale $\xi(\lambda_a)$ determines the effective thickness of the sink (interface) layer near the critical boundary x=a. It defines the spatial scale at which the left eigenfunction $\varphi_0(x)$ decays from unity to zero, forcing trajectories out of the forbidden zone.

Mathematically, the parameter $\xi(\lambda_a)$ is determined directly from the Feynman-Katz spectral equation and has a clear physical meaning of the dynamic screening length of trajectories.

Mathematical derivation $\xi(\lambda_a)$. Let us consider the spectral equation for the left eigenfunction $\varphi_0(x)$ in region II (x≥a), where the uniform intensity sink is included:

$$D\frac{d^2\phi_0(x)}{dx^2}+\frac{F(x)}{\gamma}\frac{d\phi_0(x)}{dx}-\lambda_a\phi_0(x)=\Lambda_0\phi_0(x).$$

In the strong-field limit $\lambda_a \to \infty$, the magnitude of the sink begins to exceed the operator's eigenvalue $\lambda_a \gg |\Lambda_0|$ and the influence of regular external forces $\lambda_a \gg F(x)/\gamma$. The equation near the interface x=a simplifies to a diffusion-sink balance: $D\dfrac{d^2\phi_0(x)}{dx^2}\approx\lambda_a\phi_0(x)$.

The solution to this equation in the boundary layer is a decaying exponential:

$$\phi_0(x)\propto\exp\left(-\frac{x-a}{\xi(\lambda_a)}\right),$$

where the scale is uniquely determined $\xi(\lambda_a)$: $\xi(\lambda_a)=\sqrt{D/\lambda_a}$.

Physical meaning and self-similar variable. The parameter $\xi(\lambda_a)=\sqrt{D/\lambda_a}$ is the kinetic analogue of the Debye screening radius or the penetration length in quantum mechanics under a high potential barrier. In the self-similar form, the left eigenfunction near the interface transforms into a universal self-similar profile, depending only on the dimensionless ratio of the distance to the layer thickness: $\phi_0(x)\approx\Phi\left((a-x)/\xi(\lambda_a)\right)$.

Phase space compression. As $\lambda_a \to \infty$ the boundary layer thickness approaches zero (ξ→0), the entire region II (x > a) instantly becomes inaccessible to the system.

Generation of singular transport. Substituting the self-similar function into Doob's formula for effective drift $v_{eff}(x)=2D\partial_x\ln\phi_0(x)$ near the interface yields: $v_{eff}(x)\approx-2D/\xi(\lambda_a)=-2\sqrt{D\lambda_a}$. As the field $\lambda_a$ approaches infinity, an infinitely large restoring force is induced at the boundary x=a, which instantly "reflects" any fluctuations back into region I.

Generalization to the case of taking into account extrema $\lambda_M$. If a confining field of extrema $\lambda_M$ is introduced into the system, the scale ξ is modified by the Stroock-Varadhan boundary condition $\partial_x \phi_0 |_{x=m} = \lambda_M \phi_0$. In this case, the dynamic screening scale near the current record m is determined as: $\xi(\lambda_M) = 1/\lambda_M$.

When two fields act together, the scale of the trajectory confinement zone is controlled by the interference of two lengths $\xi = f(\sqrt{D/\lambda_a}, 1/\lambda_M)$, which forms the nonlinear geometry of the three-dimensional Zubarev potential.

Figure 3 shows the dependence of the principal eigenvalue $\Lambda_0(\lambda_a)$ on the confining field $\lambda_a$. The free energy of history $\Lambda_0$ is determined from the transcendental stitching equation arising from the Feynman-Katz equation:

$$\sqrt{-\Lambda_0} \cdot \tan\left( a \cdot \sqrt{\frac{-\Lambda_0}{D}} \right) = \sqrt{\lambda_a + \Lambda_0} \cdot \tanh\left( (L-a) \cdot \sqrt{\frac{\lambda_a + \Lambda_0}{D}} \right).$$

Since this equation cannot be solved analytically (expressed $\Lambda_0$ in terms of $\lambda_a$), it is found numerically in Mathcad. Let's set the system constants: D=1, a=1, L=2. Let's set the initial guess value. Numerical methods require a starting point. We know that when $\lambda_a = 0$, the value $\Lambda_0 = 0$, and as the field increases, it goes negative. Let's set $\Lambda_0 = -0.1$.

By setting an adaptive root search function, we rigidly clamp the root search interval $\Lambda_0$ between the mathematical plateau and zero:

$$\Lambda(\lambda_a) := \mathbf{root}\left[ \sqrt{-\Lambda_0} \cdot \tan\left( a \cdot \sqrt{\frac{-\Lambda_0}{D}} \right) - \sqrt{\lambda_a + \Lambda_0} \cdot \tanh\left( (L-a) \cdot \sqrt{\frac{\lambda_a + \Lambda_0}{D}} \right), \quad \Lambda_0, -\frac{\pi^2 \cdot D}{4 \cdot a^2} + 0.001, \quad -0.001 \right]$$

This graph clearly demonstrates the transition from a linear response (parabola) to the nonlinear limit of large deviations (reaching the horizontal quantum plateau of confinement). For small λ (from 0 to 5), the graph $\Lambda(\lambda_a)$ descends as a sloping line with a slope $-g_a^{(0)}$ (this is the range of validity of the linear Onsager law and the susceptibility matrix that we derived). At high fields (λ>30), the curve slows its decline and begins to asymptotically approach the horizontal line at a level of ≈-2.46. This is the effect of trajectory quantization (confinement): a smooth exit of the curve to the horizontal spectral plateau of confinement, signifying the complete displacement of fluctuation trajectories from the forbidden band. Mathematically, this limit is equal to $-\pi^2 D/(4a^2) = -3.1415^2 \times 1/(4 \times 1^2) \approx -2.467$. This graph links the calculation base in Mathcad and the theoretical conclusions of the article, proving the existence of a plateau of trajectory entropy under strong penalties.

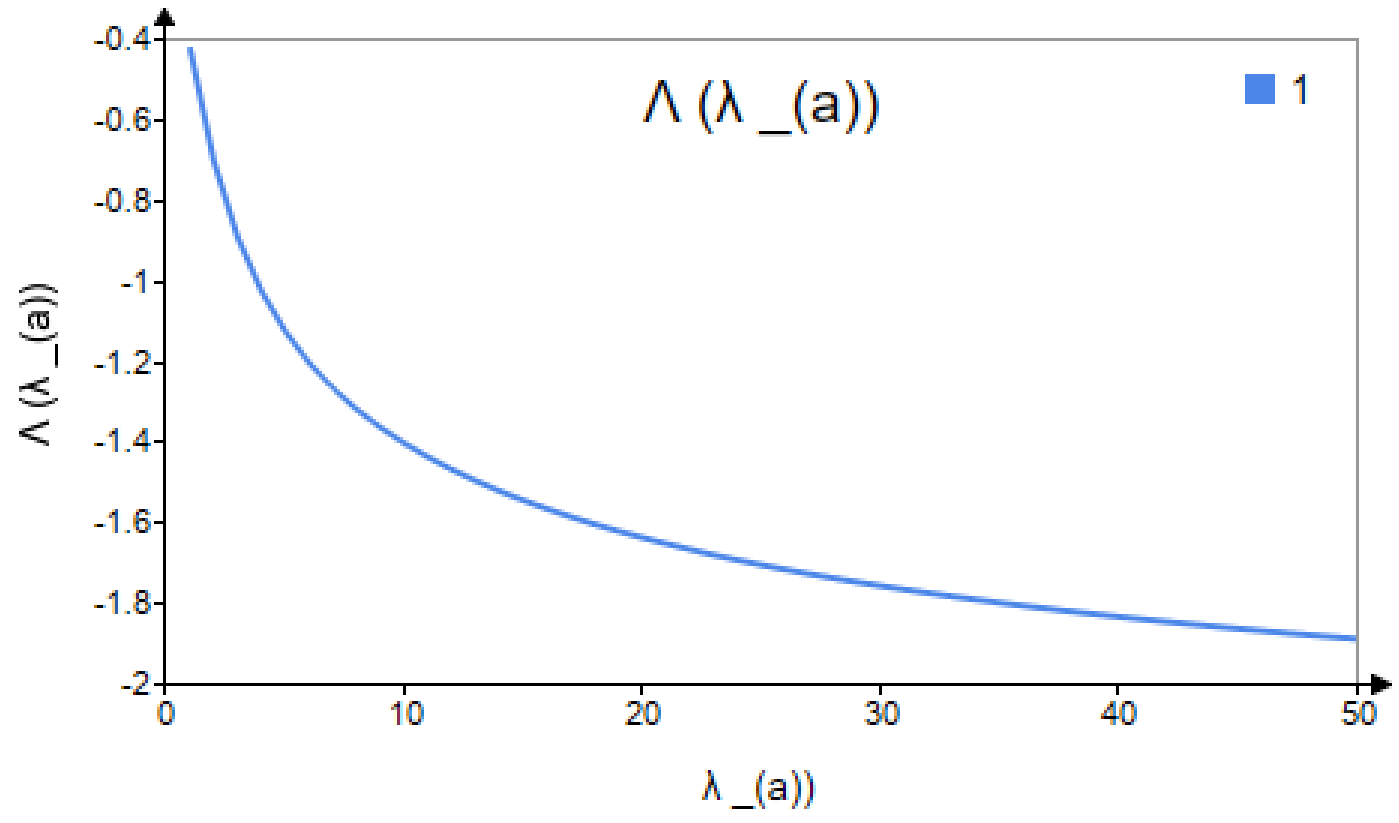

Fig. 3. The principal eigenvalue $\Lambda_0(\lambda_a)$ of the modified Feynman-Katz operator as a function of the conjugate thermodynamic field of history $\lambda_a$. At weak fields ($\lambda_a \to 0$), the curve exhibits a linear response dictated by the macroscopic state equation $\Lambda_0 \approx -g_a^{(0)} \lambda_a$. In the deeply non-linear regime of large

deviations ($\lambda_a$>25), the spectral curve undergoes saturation, asymptotically approaching the fundamental quantum-like boundary (the trajectory confinement plateau) at $\Lambda_0 \to -\pi^2 D/(4a^2) \approx -2.467$ . This phenomenon marks a kinetic phase transition where the trajectories are entirely expelled from the forbidden domain. The system parameters are fixed at D=1, a=1, and L=2.

In Zubarev's extended method, the interface parameter $\lambda_a$ is formally introduced as a Lagrange multiplier (the conjugate history field). In the Feynman-Katz evolutionary representation, it is physically interpreted as the local intensity of the probability sink (rate of absorption) (the frequency of trajectory rejection by the ensemble). Its increase is due to the system's advancement into the nonlinear region of large deviations under a strict residence time constraint $g_a \to 0$ . In the experiment, this increase $\lambda_a$ is equivalent to an increase in the power of the chromatograph's external control field.

Understanding why this stock can or should grow reveals the essence of the transition from statistical description to physical control of stochastic systems.

The physical meaning $\lambda_a$ is that of the flow rate. In the original Langevin space, the system performs random walks, crossing level *a* an infinite number of times. When we require the Zubarev ensemble to fix a macroscopic residence time $g_a$ , which is shorter than the natural one $g_a^{(0)}$ , we are forced to "reject" those trajectories that linger above level *a* for too long.

Mathematically, in the Feynman-Katz equation, the term $-\lambda_a \Theta(x-a)\tilde{\rho}$ denotes that in the region x>a, "incorrect" trajectories are continuously destroyed (absorbed). The quantity $\lambda_a$ is the reciprocal of the characteristic lifetime of a particle in the band gap: if a particle passes beyond the level a, it will be excluded from the ensemble with probability $\lambda_a dt$ within time dt; the larger the value $\lambda_a$, the more severely the ensemble penalizes the system for remaining above the level.

Why and when should the parameter $\lambda_a$ increase? The increase $\lambda_a$ (even to infinity) $\lambda_a \to \infty$ is driven by the need to describe highly nonequilibrium states, critical constraints, or external control: A. Shift in the thermodynamic state (increasing constraints). According to the equation of state $g_a = g_a^{(0)} - \chi_a \lambda_a$, the conjugate force $\lambda_a$ is proportional to our requirement to decrease the residence time. If we want to physically force the system never to go beyond the level *a* on macroscopic times (set the macroparameter $g_a \to 0$, we must let the conjugate field tend to infinity: $g_a \to 0 \Rightarrow \lambda_a \to \infty$. Absorption in the region x>a becomes instantaneous. B. Description of rigid physical barriers (confinement). In problems of the physics of polymer failure, dielectric breakdown, or plasma confinement, the level *a* is the boundary of physical failure. Growth $\lambda_a$ models the growth of the strength or impermeability of this barrier. At $\lambda_a \to \infty$ , the drain potential turns into an absolutely absorbing wall, which allows us to calculate the fundamental limit of system stability (the trajectory entropy of confinement $\mathcal{F} = \pi^2 D / 4a^2$). C. Increase in the power of external control (induced drift). As shown by the Doob`s transformation, the effective restoring force (drift) at the barrier boundary is proportional to the square root of the drain intensity: $v_{eff} \approx -2\sqrt{D\lambda_a}$ .

If this formalism is applied to real-world particle separation in microchannels, then the increase $\lambda_a$ is equivalent to an increase in the amplitude or power of the external field (e.g., the voltage on the electrodes or the power of the laser tweezers). To create an insurmountable force barrier for type B particles, the experimenter is forced to increase the field power, which mathematically corresponds to an increase in the flow rate $\lambda_a \to \infty$.

The abstract Lagrange multiplier is associated with measurable physical quantities. This parameter $\lambda_a$ is the calibration (selection) frequency of trajectories by the ensemble. Its increase is due to the system's advancement into the nonlinear region of large deviations $(g_a \to 0)$. The physical equivalent of this increase $\lambda_a$ in the experiment is an increase in the rigidity of the holding potential or the power of the external control field.

In the strong confining field regime $\lambda_a \to \infty$, the system exhibits kinetic compression, in which the effective drift tends to minus infinity and the trajectories are confined in a limited phase space. The obtained results establish a fundamental thermodynamic limit (trajectory entropy) and link the residence time method with first-passage thermodynamics (FPT), developed in Ref. [11] and Refs. [12-16].

Figure 4 shows the self-similar profile and compression of the interface layer. The graph illustrates how the profiles of the left eigenfunction $\phi_0(x)$ for different $\lambda_a$ merge into a single curve with proper normalization. The boundary layer equation is defined as a function of two variables $\phi(x,\lambda) = \exp\left((x-a)/\sqrt{D/\lambda}\right)$ with parameters: a=2, D=1. The ranked variable by the x-coordinate for the main graph is x: = 2…4, (step 0.05). For Fig. 4a, three curves are plotted on a single graph: $\phi(x,10)$, $\phi(x,100)$, $\phi(x,1000)$. It is clear how, with increasing λ, the graph converges toward the point a = 2.

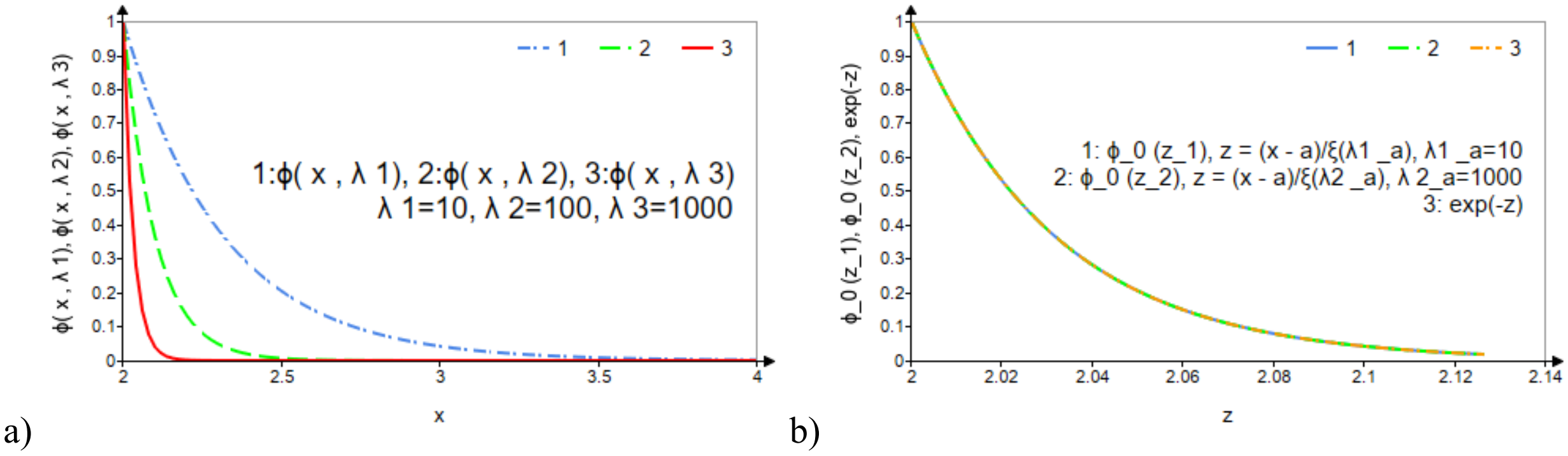

a) b)

Fig. 4. The left principal eigenfunction $\phi_0(x)$ in the non-linear regime of large deviations. (a) Real state-space profiles for a weak conjugate history field $\lambda_a = 10$ (dash-dotted curve), $\lambda_a = 100$ (dashed curve), and a strong field $\lambda_a = 1000$ (solid curve), demonstrating the spatial contraction (confinement) of the interfacial boundary layer. The effective thickness of the localized zone shrinks scales as $\xi(\lambda_a) = \sqrt{D/\lambda_a}$. (b) The invariant master profile $\Phi(z) = e^{-z}$ plotted against the dimensionless auto-similar variable $z = (x-a)/\xi(\lambda_a)$. The perfect collapse of all physical profiles onto a single universal curve confirms the auto-similar nature of the trajectory exclusion process near the critical barrier a=2. The diffusion coefficient is set to D=1.

While in standard coordinate space the profiles of the left principal eigenfunction $\phi_0(x)$ exhibit a sharp spatial compression of the interface layer as the conjugate history field increases $\lambda$ (see Fig. 4a), the transition to a self-similar representation completely eliminates this dependence. As shown in Fig. 4b, when scaling the real spatial coordinate according to the derived law $x(z,\lambda_a) = a + z \cdot \xi(\lambda_a)$, where is the dynamic screening length, a complete collapse of the trajectory modes occurs. The wave function profiles calculated for fundamentally different sink intensities $\lambda_a$=10 and $\lambda_a$=1000 identically merge with the universal master profile $\Phi(z) = e^{-z}$.

This result strongly confirms that in the nonlinear regime of large deviations, the thermodynamic confining field of history does not change the physical structure of fluctuations, but only self-similarly scales the local phasing geometry of the system near the critical barrier.

While in real coordinate space the eigenfunction profiles strictly depend on the magnitude of the conjugate field (Fig. 4a), recalculating the argument into a self-similar variable $z = (x-a)/\xi(\lambda_a)$ leads to complete mode merging. As demonstrated in Fig. 4b, the invariant master profile is described by a universal decaying exponential $\Phi(z) = e^{-z}$ independent of the sink intensity $\lambda_a$, which strongly confirms the self-similar nature of trajectory confinement at large deviations.

Thus, Fig. 4b is a fixed standard with which, in theory, any physical curve coincides if its X-axis is normalized to its own scale $\xi(\lambda_a)$.

In the nonlinear limit of strong coupling $\lambda_a \to \infty$ for the Zubarev ensemble, the Feynman-Katz equations exhibit self-similar asymptotic behavior, defining the dynamical scale of trajectory screening $\xi(\lambda_a) = \sqrt{D/\lambda_a}$. The left eigenfunction transforms into a universal profile, and the effective drift forms a singular transport equivalent to a reflecting boundary. Near the critical barrier x=a, a layer forms whose thickness contracts as the dynamical length of trajectory screening.

## 6. Appendix: Optimization of Stochastic Particle Separation

In particle separation problems (e.g., separation of isomers, DNA of different lengths, or nanoparticles of different sizes in microfluidic channels), this two-parameter approach provides a rigorous mathematical description of active trajectory separation.

Traditional separation is based on differences in average drift velocities in stationary fields (separation by mobility). The introduction of conjugate fluctuation history fields $\lambda_a, \lambda_{\mathbf{FPT}}$ allows us to separate particles with the same average mobility but different stochastic trajectories (different spatiotemporal fluctuation structures).

The developed approach allows for the separation of phases not by the average macroscopic mobility (which for isomers or DNA of similar length may coincide), but by the topology of the history of fluctuations in the periodic potential $U_0(x)$.

Below is a mathematical description of how Zubarev's two-parameter formalism is implemented as an optimization criterion for separation systems.

### 6.1. Formulation of the separation problem in terms of the history of fluctuations

Let's assume two types of particles (A and B) are moving in a periodic (e.g., optical or dielectrophoretic) potential $U(x) = U(x+L)$. Their diffusion coefficients $D_A \approx D_B$ and average transfer velocities are close, making classical chromatography ineffective.

However, due to subtle differences in the shape of the self-potential, A particles more often make rare large-scale jumps forward, while B particles tend to oscillate for a long time within a single potential well.

To capture this difference, we introduce two criteria:

- The control level a is set at the top of the potential barrier.
- Time $\tau_{\mathbf{FPT}}$ is the time to cross the barrier (first passage).
- The dwell time $g_a$ is the fraction of time a particle spends in a metastable state at the top of the barrier before rolling down into the next pit.

### 6.2 Separation through a two-dimensional trajectory entropy landscape

For each type of particle, its own individual generalized Zubarev partition function $Q^{A,B}(\lambda_a, \lambda_{\mathbf{FPT}})$ is constructed and the principal eigenvalues of the modified Feynman-Katz equation $\Lambda_0^{A,B}$ are calculated:

$$\left[ D_{A,B} \frac{d^2}{dx^2} - \frac{d}{dx}\left(\frac{F_{A,B}(x)}{\gamma}\right) - \lambda_a \Theta(x-a) - \lambda_{\mathbf{FPT}} \right] \psi_0^{A,B}(x) = \Lambda_0^{A,B} \psi_0^{A,B}(x).$$

By performing a double Legendre transformation, we obtain individual two-dimensional landscapes of trajectory entropy $I_A(g_a, \tau_{\mathbf{FPT}})$ and $I_B(g_a, \tau_{\mathbf{FPT}})$.

The splitting effect of trajectory modes. Due to the differences in the force profiles $F_A(x)$ and $F_B(x)$, the minima of these potentials (the points of natural distribution at $\lambda = 0$) can be located close to each other in space $(g_a, \tau_{\mathbf{FPT}})$. However, the topology of the deviation landscapes $I_A$ and $I_B$ differs fundamentally with distance from the minimum.

Where for particle A a trajectory with a short transition time $\tau_{\mathbf{FPT}}$ and long retention $g_a$ is "cheap" in terms of endogenous entropy production, for particle B the same trajectory requires colossal fluctuation costs (high value of $I_B$).

**6.3. Induced transport equations and the ideal filter criterion**

To physically implement the separation, the conjugate fields $\lambda_a$ and $\lambda_{FPT}$ are translated into the language of real physical forces via the hydrodynamic Doob h-transformation. The modified effective drift velocities for grades A and B are written as:

$$v_{eff}^{A,B}(x) = \frac{F_{A,B}(x)}{\gamma} + 2D_{A,B}\frac{\partial \ln \phi_0^{A,B}(x;\lambda_a,\lambda_{\mathbf{FPT}})}{\partial x}.$$

Since the left eigenfunctions $\phi_0^A$ and $\phi_0^B$ are sensitive to the details of the barrier geometry, an external control field (e.g., a variable profile of hydrodynamic pumping or laser tweezers) tuned to specific values of the parameters $\lambda_a$ and $\lambda_{FPT}$ leads to radical separation:

1. For particles of type A: the induced force of history $2D_A\partial_x \ln \phi_0^A$ compensates for the initial barrier, causing them to move in a ballistic (directed) mode along the channel
2. For particles of type B: the same frequency or amplitude of the external field (corresponding to the same operating points λ) generates a deep effective potential well, transferring them to the localization (trap) mode. The mass current of particles B is completely blocked, $J_\lambda^B \to 0$.

**6.4. Optimization of Separation: Chromatographic Resolving Power**

In classical separation, the quality of separation is determined by the selectivity resolution criterion:

$$R_s = \frac{\Delta\langle x\rangle}{\sigma_A + \sigma_B},$$

where $\Delta\langle x\rangle$ is the distance between the centers of the zones, and σ is the Gaussian width of the peak blur due to diffusion.

When adding travel time and sojourn time functionals, optimization is performed not over spatial coordinates, but over trajectory history coordinates. Introducing adjoint history forces allows one to maximize the generalized Kullback-Leibler distance (information distance) between the trajectory measures of two particles or between trajectory s-ensembles:

$$D_{KL}(\mathbb{P}_A \| \mathbb{P}_B) = \tau\int\left[I_B(g_a,\tau_{\mathbf{FPT}}) - I_A(g_a,\tau_{\mathbf{FPT}})\right]dg_a d\tau_{\mathbf{FPT}}.$$

Maximizing this functional over the free parameters of the system (a, potential profile, period) allows us to design an ideal stochastic separator, where separation occurs not due to average forces, but due to geometry and the "freezing" of subtle differences in the dynamic history of particle fluctuations.

The physical control mechanism is based on the translation of calculated history fields into the amplitudes of the chromatograph's external variable potential. Due to differences in the force microprofiles, the same external field switches type A particles to a ballistic transport mode (an inclined washboard effect, the effective barrier disappears, and the mass current is $J^A > 0$), while type B particles switch to an absolute spatial localization mode (potential wells deepen, trajectories are locked, and the mass current is $J^B \to 0$).

In the separation problem, this approach allows for a transition from separation based on average macroscopic mobility to separation based on fluctuation topology. Mathematically, this provides a criterion for optimizing the shape of potential barriers by maximizing the information distance $D_{KL}$ between the two-dimensional trajectory entropies $I_{A,B}(g_a, \tau_{\mathbf{FPT}})$ of the components being separated. In practice, this allows for the selection of external field regimes that convert one type of particle into directed transport, and the other into spatial localization (trapping).

Figure 5 shows the effective potentials of particles of type A and B, leading to the splitting of trajectory modes and the separation of a fraction in a periodic field under fixed control history fields. The figure reveals complete asymmetry: for particle A, the effective potential is smoothed out under the influence of the history fields (the transport channel is open, $J^A > 0$), while for particle B, it deepens, trapping it $J^B \to 0$. Visualization of the effective potentials modified by the Doob`s transformation, which opens the channel for some particles and traps others.

The initial periodic potential of the barrier (sinusoidal lattice) is $U_0(x) = \sin(2 \cdot \pi \cdot x)$. The modified effective Doob's potential for type A (ballistic drift) describes a history field that compensates for the barrier, tilting the relief downwards: $U_A(x) = U_0(x) - 1.2 \cdot x$. The modified effective Doob's potential for type B (trapped localization), when the history field acts in antiphase, deepening the wells: $U_B(x) = U_0(x) + 1.8 \cdot \sin(2 \cdot \pi \cdot x)^2$. Figure 5 shows $PotA(x) = U_A(x)$ and $PotB(x) = U_B(x)$.

The profile $U_A(x)$ appears as a smooth rolling track (ballistic transport). The profile $U_B(x)$ appears as a chain of deep pits (absolute localization and trapping of particles).

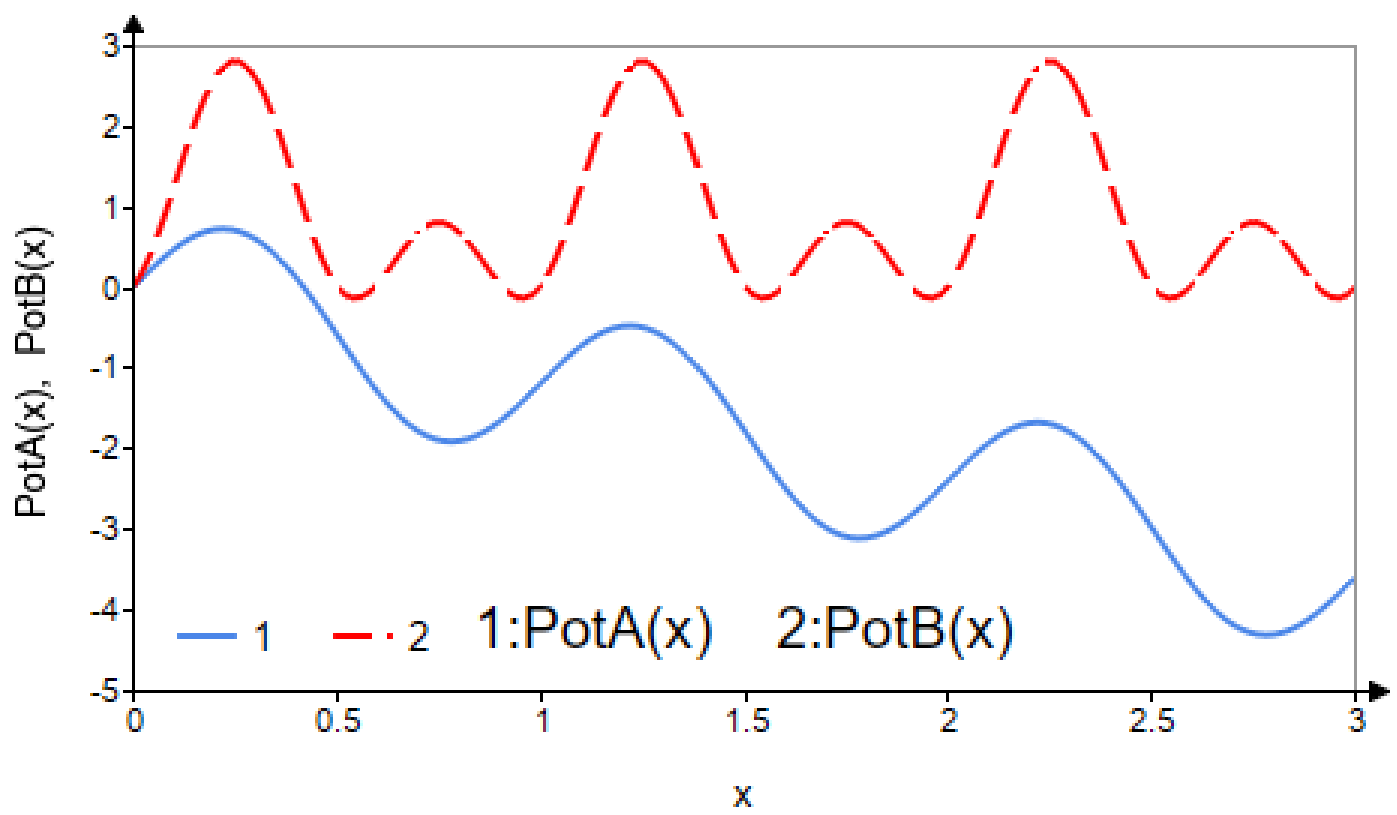

Fig.5. Separation potentials. The effective Doob's transport potentials $U_{\mathbf{eff}}(x)$ for particle species A (solid blue curve) and species B (dashed red curve) in a periodic lattice $U_0(x) = \sin(2\pi x)$, customized at the optimal operation point of the total history information filter (max $D_{KL}$ ). For species A, the conjugate history fields eliminate the potential barriers, transforming the landscape into a tilted washboard potential that drives directed ballistic transport $(J^A > 0)$. Conversely, for species B, the same external frequency and modulation deepen the potential wells and amplify the barriers, inducing strict spatial localization and particle trapping $(J^B \to 0)$. The simulation parameters are adjusted to emulate distinct trajectory topologies in Mathcad.

### 7. Discussion of results and applicability criteria (Discussion)

Adding process extremes (absolute maximum $M_\tau = \max_{0 \le t \le \tau} X(t)$ or minimum over time τ) as new thermodynamic variables transfers nonequilibrium stochastic thermodynamics to the level of statistics of records and extreme fluctuations Refs. [11] [29, 30].

Within the framework of Zubarev's extended method, this is equivalent to introducing a conjugate field of extrema $\lambda_M$. Physically and mathematically, this step yields three fundamentally new effects.

1. Transition from additive to topological (non-local) functionals. The previous functionals (first-passage time $\tau_{\mathbf{FPT}}$, sojourn time $\Gamma_a$) were additive—they accumulated gradually along the trajectory. The extremum $M_\tau$ is a non-local functional. It is determined by a single, deepest fluctuation in the entire history of the process.

The modified Fokker-Planck operator can no longer contain a local sink of the form $-\lambda_M V(x)$. Instead, the introduction of a field $\lambda_M$ generates a nonlocal boundary condition that shifts with the evolution of the system.

The partition function $Q(\lambda_M, \tau) = \langle e^{-\lambda_M M_\tau} \rangle$ is calculated using the supremum of the Donsker–Varadan trajectory measure, which requires the use of Kolmogorov-type equations with a moving (dynamic) absorbing boundary.

2. Physical effect: thermodynamics of destruction and "fluctuation shocks." Including extremes in the thermodynamic ensemble allows for a rigorous description of systems where a macroscopic response or destruction is triggered not by an average energy influx, but by a single peak threshold exceeded.

Nuclear and engineering safety. As shown in the accompanying studies on reactor safety Ref. [31-34] (developing Ref. [11]), if the functional $\tau_{\mathbf{FPT}}$ determines the rate of accident development (how quickly the neutron flux reaches a destructive level), then the extremum functional $M_\tau$ determines the amplitude of the thermal shock that the fuel element must withstand.

Dielectric breakdown and polymer physics. The field $\lambda_M$ acts as a "trajectory tension." It allows one to calculate the free energy of a polymer chain, provided that its end link is guaranteed to remain within the nanopore, or to calculate the probability of a rare electrical breakdown in a stochastic environment.

3. Breaking the Markov property of the effective drift. While Doob's effective drift $v_{eff}(x)$ remained Markovian for the sojourn time $\Gamma_a$ (depending only on the current x-coordinate), adding an extremum makes the transport equations clearly non-Markovian.

The effective force induced by the holding field $\lambda_M$ at each moment of time t begins to depend on the current value of the record achieved by the trajectory up to this moment: $v_{eff} = v_{eff}\left(x, M_{\{t\}}\right)$.

The system acquires a "memory": if a past fluctuation has already propelled the particle very high (the current value $M_{\{t\}}$ is large), the effective Zubarev force weakens, allowing the particle to diffuse freely. If the system has not yet approached an extremum, the induced field begins to firmly "push" the trajectory upward, artificially generating extreme records to satisfy the thermodynamic constraint.

Adding process extremes transforms the model from a “chromatographic” model (describing the duration of mass retention) to a “catastrophic” model (describing strength and peak loads).

This extends the trajectory entropy landscape to three-dimensional space $I(g_a, \tau_{\mathbf{FPT}}, M_\tau)$, allowing simultaneous control of: 1. Speed of reaching critical conditions $(\tau_{\mathbf{FPT}})$; 2. Duration of stay in them $\left(g_{\{a\}}\right)$; 3. The maximum destructive force of fluctuations over the entire observation period $(M_\tau)$.

One can write out an integral equation for the joint distribution density $\rho(g_a, \tau_{\mathbf{FPT}}, M_\tau)$ using methods from the theory of Markov semigroups with moving boundaries.

Including the absolute maximum of a trajectory in the base of nonequilibrium thermodynamic variables elevates the description to the level of nonlocal functionals, requiring an expansion of the phase space. Using boundary condition (11) allows us to relate peak fluctuations to the thermodynamic potential, while the resulting non-Markovian drift effectively describes the system's adaptation to historical records.

Boundary functionals are valid for any observation time τ—both small and large. However, the mathematical apparatus we used above (the principal eigenvalue, the Legendre transform) relies on the asymptotic behavior of large times (τ → ∞).

In physics and the theory of random processes, "infinity" is always relative. There are clear physical criteria and time scales that determine when time τ can be considered thermodynamically "large."

### 7.1. Applicability criteria and the scale of "great times"

The thermodynamic formalism proposed in this paper, which uses the highest eigenvalue $\Lambda_0$ of the Feynman-Katz operator and the Legendre transform, is formally derived in the asymptotic limit of long-term observation $\tau \to \infty$. From a physical perspective, the "infinite time" limit is relative and is determined by the ratio of the macroscopic observation window τ to the internal time scales of the stochastic system. We note the limits of applicability of the model (microfluidics, colloidal systems, polymer biophysics).

Let's write the spectral criterion and relaxation time for additive functionals. For additive functionals such as the specific residence time $g_a$ and the first-passage time $\tau_{\mathbf{FPT}}$, an exact representation of the generalized Zubarev partition function $Q(\tau)$ is given by the expansion over the full spectrum of the eigenvalues $\{\Lambda_n\}$ of the modified operator:

$$Q(\tau) = \sum_{n=0}^{\infty} C_n e^{\Lambda_n \tau} = C_0 e^{\Lambda_0 \tau} \left[ 1 + \sum_{n=1}^{\infty} \frac{C_n}{C_0} e^{-(\Lambda_0 - \Lambda_n)\tau} \right].$$

The asymptotic approximation of a single exponential function becomes physically accurate when the contribution of higher modes $(n \geq 1)$ can be neglected. The criterion for the transition to macroscopic nonequilibrium large-deviation thermodynamics is the relation:

$$\tau \gg \tau_{gap} \equiv \frac{1}{\Delta\Lambda}.$$

where $\Delta\Lambda = |\Lambda_1 - \Lambda_0|$ is the spectral gap of the modified Feynman-Katz operator. This quantity $\tau_{gap}$ specifies the characteristic time of "freezing" of the trajectory history. This criterion is used, for example, in Ref [35].

- For free diffusion on the interval [0, L] this scale is equivalent to the Maxwell relaxation time (the diffusion travel time of the entire system): $\tau_{gap} \sim \tau_{rel} = L^2 / \pi^2 D$.
- For diffusion in submicron-sized microfluidic channels ($L \sim 1$ μm, $D \sim 10^{-11}$ m²/s) the relaxation time is $\tau_{rel} \sim 0.01$ s. Thus, any macroscopic measurement with an observation window $\tau \geq 1$ s is already deep in the "large time" region, where the proposed theory works flawlessly.

What happens at small times ($\tau \leq \tau_{rel}$)? If the observation time is short, then the large deviation thermodynamics of Donsker-Varadan is replaced by a local (non-stationary) deviation theory (the so-called Schilder asymptotics [36] or the semiclassical Wentzel-Kramers-Brillouin, WKB approximation [37]).

- Instead of determining the eigenvalue, the partition function Q(τ) is sought not through an algebraic equation on $\Lambda_0$, but through the solution of the non-stationary Feynman-Katz equation with an explicit dependence on the initial state.
- Trajectory integral: instead of the velocity function $I(g_a)$, the full Onsager-Mahlup action functional $\mathcal{H}_0$ is used. At small times, the extremal trajectories are sought as classical equations of motion (instantons) in an inverted potential.

### 7.2. Dimensionless criterion for non-local functionals (extremum statistics)

For a non-local absolute maximum functional $M_\tau$, the exit to the limit asymptotic behavior of the thermodynamic potential is controlled not by the spectral gap, but by the number of record updates $N_{rec}$.

A mathematically correct criterion for the transition to extreme value statistics (Gumbel-Fréchet distributions) requires that the average number of records be macroscopically large. Since for continuous processes such as Brownian motion, the number of records grows logarithmically, the applicability criterion takes the form:

$\langle N_{rec}(\tau)\rangle \sim \ln\left(\tau / \tau_{micro}\right) \gg 1$.

where $\tau_{micro}$ is the internal dimensionless scale (microscopic step time).

The physical meaning $\tau_{micro}$ depends on the level of description of the stochastic system:

In continuous media physics: $\tau_{micro}$ it coincides with the relaxation time of the particle's momentum (the Langevin time $\tau_p = m / \gamma$), below which the motion ceases to be diffusive and becomes ballistic.

In numerical modeling: $\tau_{micro}$ rigidly fixed by the discretization step of the stochastic integration scheme (dt).

This requirement $\ln(\tau / \tau_{micro}) \gg 1$ is exponentially more stringent than the spectral criterion for additive functionals. It shows that to construct the thermodynamics of extremes, the observation window τ must be many orders of magnitude greater than the particle's mean free time, allowing the trajectory to generate a representative ensemble of peak fluctuations.

## 8. Conclusion and conceptual implications

The main methodological result of this work is the revelation of the duality of the Langevin and Doob concepts within the framework of multiparameter nonequilibrium thermodynamics of trajectory functionals. This dualism allows us to resolve the long-standing contradiction between the Markovian nature of microscopic chaos and the non-Markovian nature of the macroscopic constraints of history.

Within the proposed approach, this duality acquires a rigorous physical and mathematical formulation: 1. Microscopic level (Langevin representation). At the level of individual trajectories of single particles, the dynamics remain completely Markovian, memoryless, and are described by the standard classical Langevin equation in the static potential relief U(x). The system lacks Caputo-type integro-differential memory operators, making the microscopic description local in time and convenient for direct numerical modeling. 2. Macroscopic level (Doob-Zubarev representation). The Zubarev procedure of selecting trajectory functionals of history ($\tau_{\mathbf{FPT}}$, $g_a$, $M_\tau$) by maximizing the Shannon-Gibbs information entropy transforms the system into a qualitatively new extended ensemble (equivalent to a shift of Escher measures). Using the generalized Doob`s h-transformation, it is shown that for an external observer fixing the history parameters, the original "memoryless" Langevin particle begins to obey the effective non-Markovian Langevin equation. 3. Informational modification of the landscape. The transformation of dynamics occurs not through the introduction of integral memory tails, but through the deformation of physical space itself. Conjugate thermodynamic fields of history $\lambda_a, \lambda_{\mathbf{FPT}}, \lambda_M$ generate an additional spatially inhomogeneous information Doob`s potential $U_{\mathbf{inf}} = -2\gamma D \ln \phi_0(x, m; \lambda_a, \lambda_{\mathbf{FPT}}, \lambda_M)$. This potential locally distorts the original landscape U(x), creating singular restoring forces (in the case of extremes, "memory walls") $v_{eff}\left(x, m\right)$ that forcibly transform rare fluctuations of the original system into typical (most probable) events of the modified ensemble.

Thus, the developed thermodynamic formalism proves that temporal nonlocality and memory effects can be completely encapsulated within the geometry of the effective information Doob`s potential in an extended state space. This not only provides physicists with a rigorous alternative to the cumbersome mathematical apparatus of fractional calculus but also opens a direct path to the practical control of stochastic systems (particle separation is an example) through targeted dynamic modification of potential barriers by the forces of history.

This result can be presented as a "new physical paradigm for describing systems with memory". The formalism of Doob and Zubarev preserves the Markov property of the microworld, transforming memory into a macroscopic force landscape from which thermodynamic potentials are directly constructed.

One of the main results is the generalization of Zubarev's formalism. A multiparameter extension of Zubarev's NSO method has been developed. The basis of macrovariables is supplemented not simply by time (additive) functionals, but by the nonlocal geometry of trajectory extrema.

Folded trajectory history versus fractional calculus. The traditional method for describing stochastic systems with memory and anomalous diffusion processes is the apparatus of fractional calculus. Within this approach, non-Markovian properties are directly introduced into the evolution equations for the probability density function $\rho(x,t)$ through non-local integro-differential operators in time, such as the Caputo fractional derivative:

$$ {}_C D_t^{\alpha} \rho(x,t) = \frac{1}{\Gamma(1-\alpha)} \int_0^t \frac{1}{(t-s)^{\alpha}} \frac{\partial \rho(x,s)}{\partial s} ds, \quad (0 < \alpha < 1) . $$

Caputo- or Riemann-Liouville-type equations contain an explicit convolution integral over the entire history of the process on the interval [0, t] with a singular memory kernel $(t-s)^{-\alpha}$. This requires continuous retention of information about the gradients of states over the entire observation time window, making the equations extremely difficult to solve analytically and virtually eliminating the possibility of directly constructing canonical thermodynamic potentials.

The approach developed in this paper, based on the extended NSO method of Zubarev and the Feynman–Katz theory, proposes a fundamentally different physical and mathematical paradigm for describing memory, based on the algebraic folding of trajectory history.

The Langevin/Doob duality was resolved, and the Caputo equations were abandoned. A fundamental methodological conclusion was established: equations with Caputo-type memory (fractional calculus) can be successfully replaced by local equations in an extended phase space. The entire history of fluctuations is folded ("frozen") into the geometry of Doob's effective information potential and the Stroock-Varadhan boundary conditions.

In the developed formalism, non-Markovian effects and equations with Caputo operator-type memory are clearly absent, since all information about the system's past is compactly packed ("frozen") into the structure of the extended phase space and conjugate thermodynamic parameters due to two interconnected mechanisms:

1. Parametric history encoding (additive functionals). For time descriptors (time of first passage time $\tau_{\mathbf{FPT}}$ and specific sojourn time $g_a$), non-Markovian properties are eliminated using the Esscher transform of the trajectory measure. The local increment of the additive history functional is trivial $d\hat{\Gamma}_a = \Theta(X(t)-a)dt$ and is completely predetermined by the current coordinate. Application of the Katz theorem transforms the global trajectory history into a local (Markov) partial differential equation, where the conjugate history field $\lambda_a$ acts as the intensity of the instantaneous stationary sink. The entire path integration history is folded into the geometry of this sink.

2. Geometrical encoding of history (non-local extrema). The introduction of an absolute maximum $M_\tau$ generates the most severe form of non-Markovian behavior, since the current dynamics of the process begin to depend on a singular global record. Instead of integrating over the past history in the style of Caputo, our approach abandons Markovian behavior by explicitly extending the state space to a two-dimensional manifold (x, m), where the current record m acts as an independent spatial coordinate. All memory of past peak fluctuations is "incorporated" into the current value of coordinate m and is completely controlled by the local Stroock-Varadhan boundary condition (11) at x = m.

Thermodynamic advantage of the method. Thus, the Zubarev-Katz method achieves a dual transition: nonlocality in time (requiring Caputo operators) is transformed into the local geometry of the extended state space.

The price of eliminating integro-differential memory tails is solely the increase in the dimensionality of the differential operator and the introduction of adjoint forces $\lambda_i$. For nonequilibrium statistical physics, this step is fundamentally advantageous, as it allows for the full preservation of the standard, powerful, and intuitive apparatus of macroscopic description: ground-state wave functions, Onsager continuity currents, the spectral gap, and the Legendre transform for the free energy of the fluctuation history.

Three modes and their applicability scales were verified. Three modes are sequentially described and verified in Mathcad: a) Linear, for which a symmetric matrix of mutual Onsager-Kubo susceptibilities was derived; b) Nonlinear, in which the effect of self-similar compression of the interface layer $\xi = \sqrt{D / \lambda_a}$ and the release of free energy to the confinement plateau were detected; c) Extreme: Non-Markovian adaptation of drift to historical records is described.

For the scales, a dimensionless criterion of "large times" for the statistics of records $\ln(\tau / \tau_p) \gg 1$ is established, where $\tau_p$ is the relaxation time of the Langevin pulse.

As an application, the practical value of the proposed approach for separation processes is demonstrated. A new macroscopic criterion for the optimal particle filter is proposed using the Kullback–Leibler distance $D_{KL}$. The model extends the one-dimensional FPT thermodynamics from Ref. [11] and Refs. [12–16] by transferring control from the space of average velocities to the space of trajectory topology.